\documentclass[twocolumn]{aastex6}

\begin{document}

\title{Orbital Motion, Variability, and Masses in the T Tauri Triple System}

\author{G. H. Schaefer\altaffilmark{1}, Tracy L. Beck\altaffilmark{2}, L. Prato\altaffilmark{3}, \& M. Simon\altaffilmark{4}}

\altaffiltext{1}{The CHARA Array of Georgia State University, Mount Wilson Observatory, Mount Wilson, CA 91023, USA; schaefer@chara-array.org}
\altaffiltext{2}{Space Telescope Science Institute, 3700 San Martin Drive, Baltimore, MD 21218, USA}
\altaffiltext{3}{Lowell Observatory, 1400 West Mars Hill Road, Flagstaff, AZ 86001, USA}
\altaffiltext{4}{Department of Physics and Astronomy, Stony Brook University, Stony Brook, NY 11794, USA}

\begin{abstract}

We present results from adaptive optics imaging of the T Tauri triple system obtained at the Keck and Gemini Observatories in 2015$-$2019.  We fit the orbital motion of T Tau Sb relative to Sa and model the astrometric motion of their center of mass relative to T Tau N.  Using the distance measured by {\it Gaia}, we derived dynamical masses of $M_{\rm Sa} = 2.05 \pm 0.14$ M$_\odot$ and $M_{\rm Sb} = 0.43 \pm 0.06$ M$_\odot$.  The precision in the masses is expected to improve with continued observations that map the motion through a complete orbital period; this is particularly important as the system approaches periastron passage in 2023.  Based on published properties and recent evolutionary tracks, we estimate a mass of $\sim$ 2 M$_\odot$ for T Tau N, suggesting that T Tau N is similar in mass to T Tau Sa. Narrow-band infrared photometry shows that T Tau N remained relatively constant between late 2017 and early 2019 with an average value of $K$ = 5.54 $\pm$ 0.07 mag.  Using T Tau N to calibrate relative flux measurements since 2015, we found that T Tau Sa varied dramatically between 7.0 to 8.8 mag in the $K$-band over timescales of a few months, while T Tau Sb faded steadily from 8.5 to 11.1 mag in the $K$-band.  Over the 27 year orbital period of the T Tau S binary, both components have shown 3--4 magnitudes of variability in the $K$-band, relative to T Tau N.

\end{abstract}


\section{Introduction}

T Tauri is a young hierarchical triple system in the Taurus star forming region.  The optically dominant component T Tau North (T Tau N) is the prototype for the class of T Tauri objects \citep{joy45} and has a spectral type of K0 \citep{luhman18}.  The infrared companion, T Tau South (T Tau S), was discovered at a separation of $\sim$ 0\farcs7  \citep{dyck82} and was subsequently revealed to be a close binary with a separation of $\sim$ 0\farcs05 \citep{koresko00}.  The spectrum of T Tau Sa appears to be relatively featureless while T Tau Sb has the infrared spectrum of an early M star \citep{duchene02}.  

The orbital motion in the T Tau triple has been monitored for almost a complete period ($P \sim 27$ yr) over the past two decades \citep{kohler00, kohler08, kohler16, koresko00, duchene02, duchene05, duchene06, furlan03, beck04, mayama06, schaefer06, schaefer14, skemer08, ratzka09, csepany15}.  Although T Tau N is one of the most massive and luminous T Tauri stars known and T Tau S is undetected in the optical \citep{stapelfeldt98}, modeling the spectral energy distribution of T Tau S \citep{koresko97} and mapping the orbital motion in the triple system \citep{duchene06,kohler08,kohler16,schaefer14} suggest that T Tau Sa is at least as massive as T Tau N.

Along with the positions of the three components, high spatial resolution observations provide measurements of their relative fluxes.  According to \citet{beck04}, the near-infrared flux of T Tau N remained constant from 1994 to 2002.  The first spatially resolved observations of T Tau Sa,Sb \citep{koresko00,duchene06} occurred about a year after the last periastron passage ($T \sim 1996.1$).  At the time of the discovery, T Tau Sa was about 2 mag brighter than Sb.  During 2002$-$2007, the flux of T Tau Sa entered a highly variable phase where it ranged from $\sim$ 2 mag fainter than Sb to 0.8 mag brighter than Sb.  The variability of T Tau Sa then appeared to decrease through early 2014, while it remained fainter than Sb \citep{schaefer14}.  \citet{csepany15} and \citet{kasper16} reported that T Tau Sa was again brighter than Sb in late 2014 through 2015.

Evidence suggests that T Tau Sa is enshrouded in a small (2--3 AU), moderately opaque, edge-on disk \citep{beck04,duchene05,skemer08,manara19}.  Beck et al.\ and Duch\^ene et al.\ speculated that changes in the brightness of Sa could be caused by variable extinction, where the star light intercepts thicker and thinner portions of the circumstellar disk as it rotates around the star.  Alternatively, \citet{vanboekel10} argue that the short-term variability is caused by variable accretion.  They speculated that the enhanced variability in the early to late 1990's was induced by a tidal perturbation of the disk following periastron passage.  Plausibly, both phenomena could contribute to the system's variability. 

In this paper we present new adaptive optics (AO) measurements of the relative positions and fluxes of the components in the T Tau triple system obtained in 2015$-$2019.  Based on these data and measurements in the literature, we compute an updated orbit fit to model the motion of T Tau Sb relative to Sa, as well as the motion of their center of mass relative to T Tau N.  We derive dynamical masses of T Tau Sa and Sb from the orbital parameters.  We also present photometry of the three components sampled at weekly to yearly timescales and discuss the variability of the system.

\section{High Resolution Near-Infrared Imaging}

\subsection{Astrometry and Flux Ratios}

AO imaging provides precise measurements of the orbital motion and relative flux ratios of the three components in the T Tau system.  At the Keck Observatory, natural guide star AO observations were obtained using the NIRC2 narrow-field camera \citep{wizinowich00new} on the Keck II Telescope.  At Gemini Observatory, observations were obtained using the Altair AO system and the NIRI f/32 camera \citep{hodapp03}.  Images were recorded in the narrow-band $K$ continuum filter during every epoch and in narrow-band $H$ continuum and $L$-band emission line filters (Br$\alpha$ and PAH) during some epochs.  The $K$ and $H$-band images were flatfielded using dome flats.  Sets of dithered images were recorded and subtracted to remove the background.  In the $L$-band we created flats from the sky background in the science frames.

\begin{figure}
  \plotone{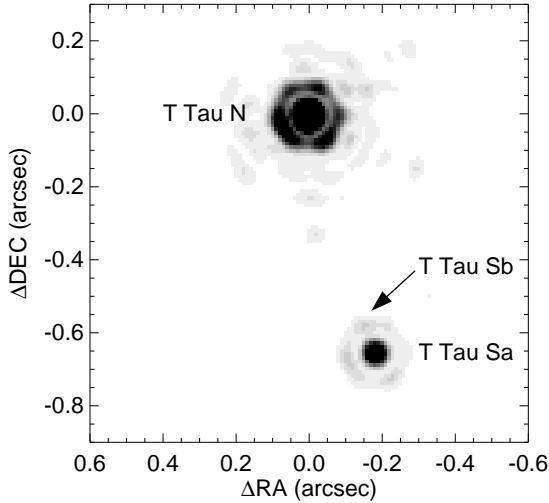}
\caption{Coadded Keck AO image of the T Tau triple in the $K$ continuum filter on UT 2019Jan20.  The positions of T Tau N, Sa, and Sb are marked.
\label{fig.image}}
\end{figure}

\begin{figure}
  \plotone{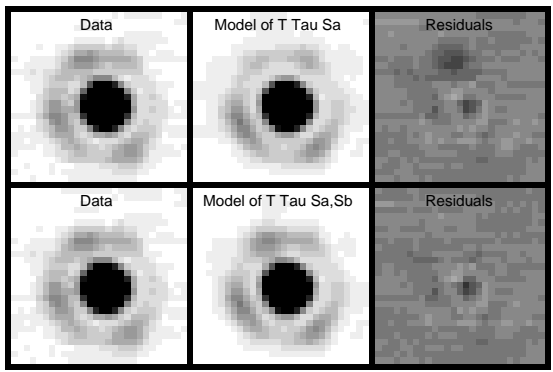}
\caption{PSF fitting of T Tau Sa,Sb for a single Keck AO image obtained in the Kcont filter on UT 2019Jan20.  The top row shows the results of modeling only T Tau Sa as a single star using T Tau N as the PSF.  The position of T Tau Sb jumps out in the residuals between the data and the model.  The bottom panel shows the results of modeling both T Tau Sa and Sb as a binary using T Tau N as the PSF.  The residuals show that T Tau Sb is fit cleanly. 
\label{fig.psffit}}
\end{figure}

T Tau N was used as a simultaneous point spread function (PSF) reference to model the position and relative flux ratios of T Tau Sa and Sb \citep[e.g.,][]{schaefer14}.  As shown in Figure~\ref{fig.image}, T Tau Sb was $\sim$ 2 mag fainter than Sa during the observations, and the position of Sb lies near the diffraction ring of Sa.  However, despite the challenge of resolving both components, Figure~\ref{fig.psffit} demonstrates that we were able to recover the position of T Tau Sb using T Tau N as a simultaneous PSF to model the close pair.  

For the Keck NIRC2 measurements, we corrected the binary positions using the geometric distortion solutions published by \citet{yelda10}, prior to the optical realignment of the AO system on 2015 April 13, and by \citet{service16} after the realignment.  For the pre-2015 observations, we used a plate scale of 9.952 $\pm$ 0.001 mas~pixel$^{-1}$ and subtracted $0\fdg252 \pm 0\fdg009$ from the raw position angles to correct for the orientation of the camera relative to true north.  After 2015 April 13, we used a plate scale of 9.971 $\pm$ 0.004 mas~pixel$^{-1}$ and subtracted $0\fdg262 \pm 0\fdg020$ from the measured position angles.  For the Gemini measurements we corrected for the radial barrel distortion\footnote{https://www.gemini.edu/instrumentation/niri/capability} and applied a plate scale of 21.9 $\pm$ 0.1 mas and field orientation of $0\fdg00$ $\pm$ $0\fdg05$.

Table~\ref{tab.sepPA} reports the Julian year, binary separation (in milliarcseconds; mas), position angle measured east of north, and flux ratios measured in each filter for each pair of components in the T Tau system.  The positions were averaged over the measurements from individual frames in the $K$ continuum band, and uncertainties were computed from the standard deviation.  During the observations, the separation of T Tau Sa,Sb was below the diffraction limit of the telescopes in the $L$-band, and the binary fit would not converge to a stable solution.  Therefore, during the analysis of the $L$-band observations, we fixed the relative separation of T Tau Sa,Sb based on the $K$-band measurements during each epoch and solved for the flux ratios.  In the $H$-band, the fluxes of T Tau Sa and Sb are very faint compared with T Tau N.  Therefore we measured the flux ratio from a coadded image of all frames, but adopted uncertainties based on the standard deviation of fits to the individual frames.

We obtained two images of T Tau using the slit-viewing camera (SCAM) for the NIRSPEC spectrograph behind the AO system on the Keck II Telescope on UT 2020 Jan 30.  We detected a component at a separation of $\sim$ 673 mas and a flux ratio of $\sim$ 0.18 relative to T Tau N, consistent with the last measured position of T Tau Sa in 2019.1.  PSF fitting did not reveal the presence of T Tau Sb.  However, in the first image, the center of T Tau N was partially saturated, and in the second, the signal-to-noise on the southern component was low, impacting our ability to detect Sb.  By adding in a fake companion at the expected location of T Tau Sb and visually inspecting the images, we suspect that Sb is at least faint as it was in 2019.0 ($\gtrsim$ 3 mag fainter than Sa). 

\label{sect.ao}

\subsection{Absolute Photometry}

On nights at the Gemini Observatory when conditions were photometric, observations were obtained of the near-infrared flux standard HD 22686.  We performed aperture photometry using the aper.pro routine in the IDL astronomy library\footnote{https://idlastro.gsfc.nasa.gov/}.  We used an aperture radius of 100 pixels in the Hcon and Kcon filters, and a smaller radius typically of 60 pixels in the Br$\alpha$ filter to minimize the number of background counts at longer wavelengths.  For T Tau, the aperture included the flux from all three components.  We centered the aperture on T Tau N at Hcon and Kcon, because the northern component dominates the light in these bands, and we centered half-way between T Tau N and S in Br$\alpha$ where the flux ratio of the northern and southern components are nearly equal.

We calibrated the total flux of T Tau by comparing with the flux measured on HD 22686.  We did not apply a correction for airmass because the targets were observed at similar airmasses ($\Delta z$ between targets ranged from 0.003 to 0.5) and the expected correction based on standard extinction curves \citep{tokunaga02} is smaller than the uncertainties in the measured values.  We used the narrow-band Hcon, Kcon, and Br$\alpha$ filters as proxies for the $HKL$ fluxes.  We computed the mean and standard deviation of the fluxes measured in the individual files and used sigma clipping to reject measurements that were more than 3\,$\sigma$ discrepant from the mean.   We calibrated the fluxes by adopting the magnitudes of $H$ = 7.186 $\pm$ 0.009 mag, $K$ = 7.186 $\pm$ 0.008 mag, and $L'$ = 7.199 $\pm$ 0.008 mag for the flux standard \citep{guetter03,leggett03}.  We then used the relative fluxes reported in Table~\ref{tab.sepPA} to partition the total flux of T Tau into the magnitudes measured for each component.  The absolute photometry is presented in Table~\ref{tab.phot}.
  
The strength and variability of emission lines in the spectra of young stars complicates the comparison of magnitudes measured between the broadband and narrowband continuum filters.  Using the narrowband filters was necessary to avoid saturation on T Tau in the Keck and Gemini AO images.  Therefore, some caution is advised when comparing the magnitudes reported here to true broadband values.  However, with the additional measurements presented in this paper, there is a growing set of relative flux measurements of the T Tau system in the narrow-band continuum filters \citep[e.g.,][]{schaefer06} that can be used to study the variability of the components over the course of the orbital period of the close pair.

\label{sect.phot}

\section{Orbital Motion in the T Tau Triple}

We fit the relative orbit of T Tau Sa,Sb to the positions in Table~\ref{tab.sepPA} and measurements in the literature \citep{kohler00, kohler08, kohler16, koresko00, duchene02, duchene05, duchene06, furlan03, beck04, mayama06, schaefer06, schaefer14, skemer08, ratzka09}.  We used a Newton-Raphson method to minimize $\chi^2$ by calculating a first-order Taylor expansion for the equations of orbital motion.  Table~\ref{tab.orb} lists the orbital parameters including the period $P$, time of periastron passage $T$, eccentricity $e$, angular semi-major axis $a$, inclination $i$, position angle of the line of nodes $\Omega$, and argument of periastron $\omega$.  For visual binary orbits, there is a $180^\circ$ ambiguity in the values of $\Omega$ and $\omega$.  This ambiguity can be resolved using radial velocity measurements to establish the direction of motion.

\begin{figure*}
\plottwo{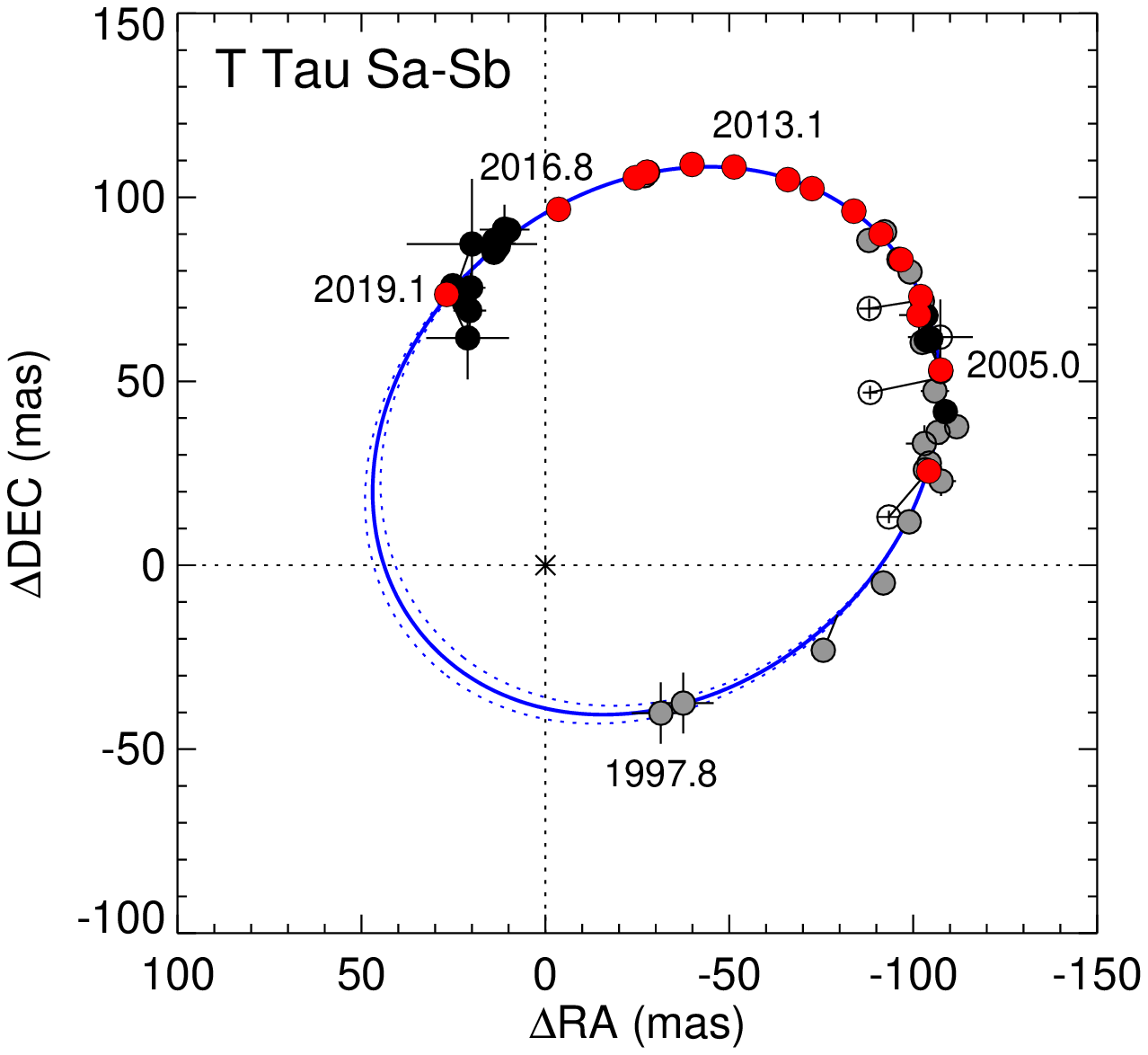}{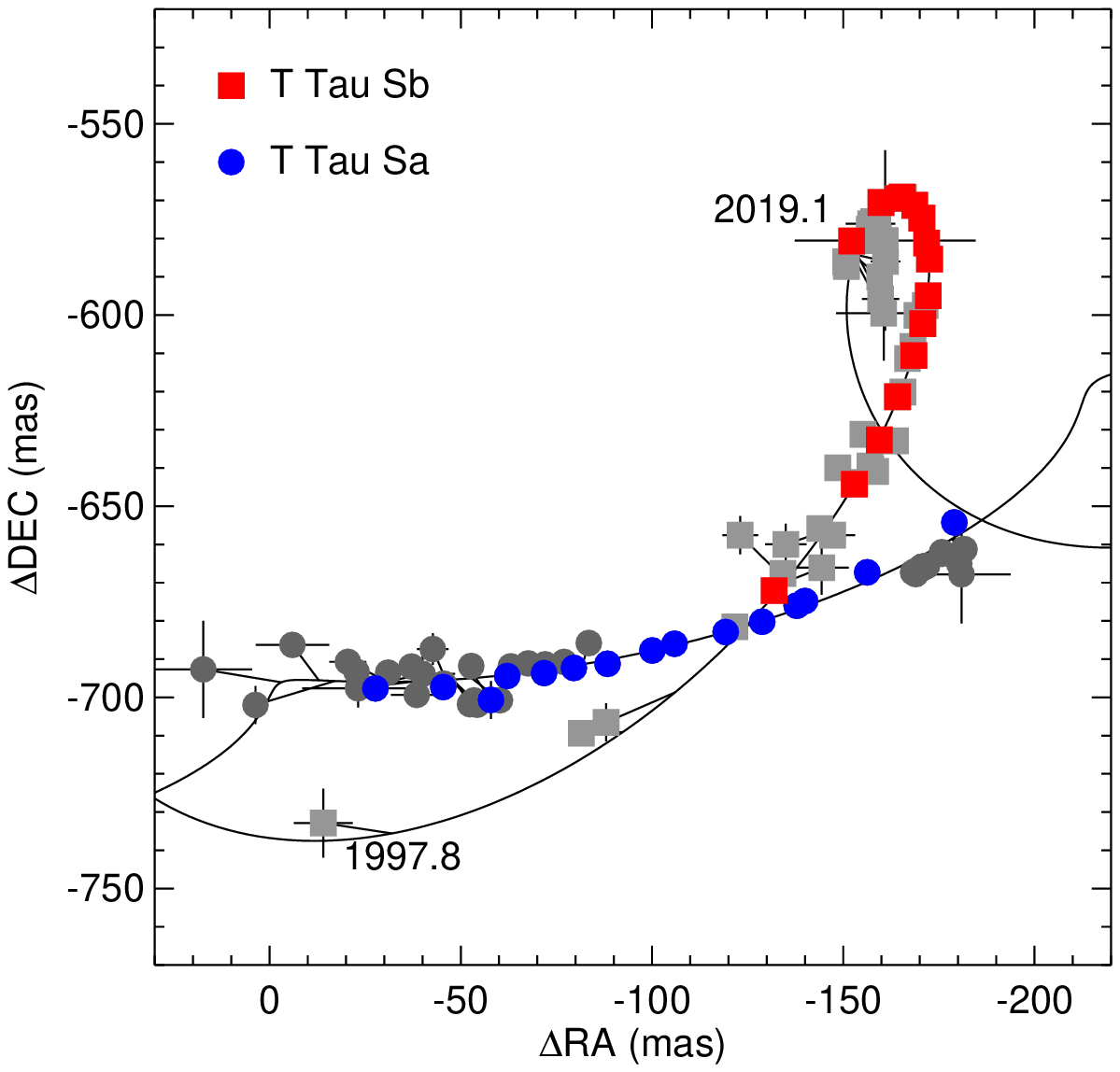}
\caption{{\it Left:} Orbital motion of T Tau Sb relative to Sa.  The red circles are Keck NIRC2 measurements presented in Table~\ref{tab.sepPA} and \citet{schaefer14}.  The black circles are Gemini NIRI measurements \citep[this work;][]{beck04,schaefer06}. The gray circles are published values from the literature \citep{kohler00, kohler08, kohler16, koresko00, duchene02, duchene05, duchene06, furlan03, beck04, mayama06, skemer08, ratzka09}. The best fit orbit and uncertainties ($P = 27.18 \pm 0.72$ yr) are overplotted in solid and dotted blue lines, respectively.  Four measurements with large residuals \citep{mayama06, skemer08, ratzka09} were not included in the fit; these are plotted as open circles. Uncertainties are shown with crosses; for much of the data these are smaller than the points themselves.  {\it Right: } Astrometric center-of-mass motion of T Tau Sa and Sb relative to T Tau N.  The blue circles (T Tau Sa) and red squares (T Tau Sb) highlight the high precision astrometry from Keck.  The dark gray circles (T Tau Sa) and light gray squares (T Tau Sb) show results from Gemini and the literature.
\label{fig.orbit}}
\end{figure*}

With all three components in the AO field of view, T Tau N serves as a reference to map the astrometric center-of-mass motion of the close pair.  The astrometric motion provides the mass ratio of the close pair, $M_{\rm Sb}/M_{\rm Sa}$.  We fit the astrometric motion by following the same approach outlined in \citet{schaefer12}.  We searched through a range of mass ratios to compute the expected location of the center of mass of T Tau Sa,Sb relative to N.  For each trial mass ratio, we fit a representative orbit to the center of mass motion of S relative to N and selected the mass ratio that minimized the $\chi^2$ between the calculated position of the center of mass and the orbit fit.  An incorrect mass ratio will produce residual reflex motion that cannot be fit by a simple Keplerian orbit.  We found a best fitting mass ratio of 0.210 $\pm$ 0.028 (Table~\ref{tab.orb}).  The relative and astrometric orbit fits are shown in Figure~\ref{fig.orbit}.   

\begin{figure}
\plotone{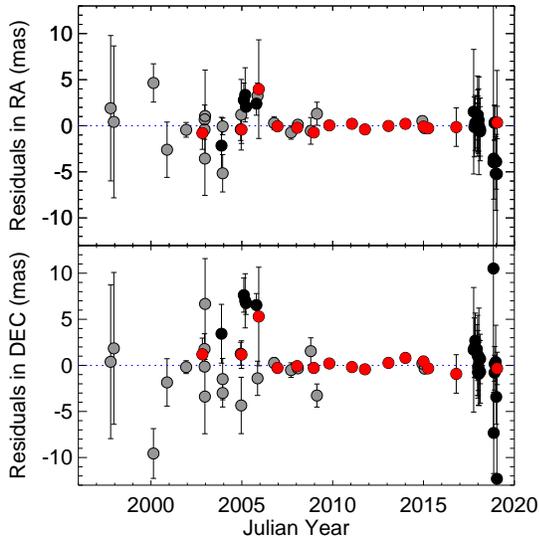}
\caption{Residuals between the measured position of T Tau Sb relative to Sa and the orbit fit.  The red circles are the Keck NIRC2 measurements, black circles are the Gemini NIRI measurements, and gray circles are from the literature.  
\label{fig.residual}}
\end{figure}

Figure~\ref{fig.residual} plots the residuals between the measured positions of T Tau Sb relative to Sa compared with the predictions from the orbit fit.  There is significant scatter in the recent Gemini observations, especially as the separation of the close pair decreases below 90~mas in 2018.8 and later.  As a check on the measured flux ratios, we fit a visual orbit for T Tau Sa,Sb to only the Keck observations and earlier measurements in the literature (excluding the Gemini observations reported here).  We then computed the expected position of T Tau Sa,Sb at the time of the Gemini observations based on this orbit fit.  Fixing the relative separation of the close Sa,Sb pair, we performed another PSF fit to the Gemini images and solved for the component flux ratios and separations relative to T Tau N.  The flux ratios derived from the constrained fit are consistent within 1\,$\sigma$ with the results reported in Table~\ref{tab.sepPA} (except for the Br$\alpha$ flux ratio in 2019.0410 which is discrepant by 1.6$\sigma$).  This provides confidence that the flux ratios are likely reliable, despite the large scatter in the Gemini positions.

\section{Dynamical Masses of T Tau S\MakeLowercase{a} and S\MakeLowercase{b}}

The relative orbit of a binary system provides a measurement of the total mass if the distance is known.  To derive masses of the components in T Tau, we used distances of 148.7 $\pm$ 1.0 pc measured from the trigonometric parallax with the Very Long Baseline Array \citep[VLBA][]{galli18} and 143.74 $\pm$ 1.22 pc derived from the {\it Gaia} Data Release 2 \citep[DR2;][]{gaia18,bailerjones18}.  These two distance measurements are discrepant by 3\,$\sigma$, producing a systematic difference in the total mass derived for T Tau Sa and Sb, as shown in Table~\ref{tab.mass}.  \citet{galli19} discusses a comparison of several other sources that have both radio and {\it Gaia} parallaxes.

The VLBA parallax is based on mapping the motion of the radio emission from T Tau Sb and must account for the orbital motion of Sb relative to Sa \citep{loinard07,galli18}.  \citet{galli18} attempted to fit an acceleration term caused by the motion relative to T Tau N, but found this contribution to be negligible.  Based on the orbit fit for the center-of-mass motion of T Tau S relative to T Tau N (see Sect.~\ref{sect.ttauN}), we find a small acceleration term of $\sim$ 0.028 mas\,yr$^{-2}$ over the time frame of the VLBA observations.  The {\it Gaia} parallax is based on the visible light from T Tau N; the measurement has a small amount of excess noise (0.12 mas).  In the subsequent discussion we opt to use the masses derived from the {\it Gaia} distance because it is less complicated by the orbital motion of the close pair.  The accuracy of the parallax should improve in the final {\it Gaia} data release.

Combining the mass ratio from the astrometric motion with the total mass from the relative orbit provides individual masses of $M_{\rm Sa} = 2.05 \pm 0.14$ M$_\odot$ and $M_{\rm Sb} = 0.43 \pm 0.06$ M$_\odot$.  Currently the masses are measured with a precision of 6.7\% and 12.7\% for T Tau Sa and Sb, respectively.  We expect the precision to improve to 2$-$5\% by continuing to map the orbital motion for a complete orbital period through the next periastron passage (expected in 2023.3).

\section{Orbit of T Tau S relative to T Tau N}
\label{sect.ttauN}

While fitting for the astrometric motion, we applied a constraint on the total system mass (N+Sa+Sb) when solving for the representative orbit of T Tau N,S.  As discussed by \citet{schaefer06}, a broad range of orbital parameters can be used to fit an orbit with limited coverage; often with a tail of eccentric solutions that yield very large masses.  The constraint on the total mass of the system does not significantly impact the final value of the mass ratio of the close pair, however, it does provide a more realistic set of orbital parameters for the wide pair that can be used to predict and back-track the expected motion in the triple system.

\begin{figure}
\plotone{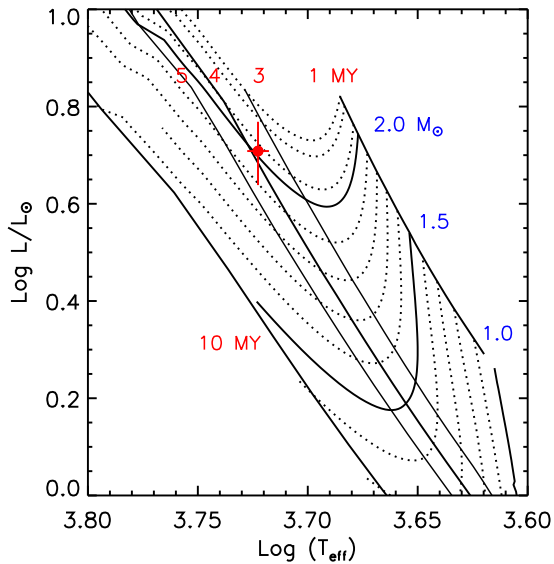}
\caption{Evolutionary models computed by \citet{feiden16} that include magnetic fields.  The mass tracks are plotted at 0.1~$M_\odot$ intervals from 1.0 to 2.0~$M_\odot$ and then at 2.05, 2.15, and 2.20~$M_\odot$.  Three of the tracks are labeled in blue and shown as solid lines for 1.0, 1.5, and 2.0~$M_\odot$.  The isochrones are labeled in red and plotted at 1, 3, 4, 5, and 10~MY.  The red circle shows the location of T Tau N (see Sect.~\ref{sect.ttauN}).
\label{fig.hrd}}
\end{figure}

We adopted the combined mass of T Tau Sa+Sb from the visual orbit and the {\it Gaia} distance ($M_{\rm Sa+Sb}$ = 2.48 $\pm$ 0.16 $M_\odot$).  We estimated the mass of T Tau N using the magnetic models of stellar evolution computed by \citet{feiden16}.  We used the luminosity derived by \citet{loinard07} scaled to the {\it Gaia} distance \citep{bailerjones18} and assumed an effective temperature of 5280 $\pm$ 60 K based on the spectral type of K0 adopted by \citet{luhman18} and the temperature scale derived by \citet{pecaut13}.  These stellar parameters correspond to a mass of $M_{\rm N}$ = 2.03 $\pm$ 0.12 $M_\odot$ and an age of  3.8 $\pm$ 0.7 MY when compared with the evolutionary tracks, as shown in Figure~\ref{fig.hrd}.

High-resolution, infrared spectra of T Tau N in the $H$-band indicate a K5 spectral type for T Tau N (e.g., R.~Lopez-Valdivia et al.\ in prep; L.~Prato in prep).  The lower effective temperature, 4200--4400 K, implied by this result may represent the impact of starspots with a large filling factor on the photospheric flux \citep{gully-santiago17}.  Discussion of the discrepancy between the K0 spectral type determined at optical wavelengths and the much later K5 type derived from infrared observations is beyond the scope of this paper and will be addressed in a forthcoming paper (L.~Prato in prep).

\begin{figure}
\plotone{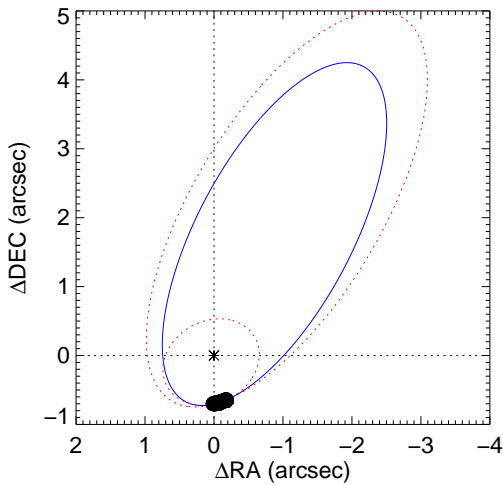}
\caption{Orbital motion of T Tau S relative to N.  The position of T Tau N is marked by the asterisk at the origin.  The black circles show the measured positions computed for the center-of-mass of T Tau Sa,Sb.  The blue line shows the best fit orbit while the red dotted lines show the range of orbital fits in Table~\ref{tab.wide}.    
\label{fig.wide}}
\end{figure}

When applying the constraint on the wide orbital motion, we limited the total system mass of the three components to be within 4.51 $\pm$ 0.59 $M_\odot$.  The uncertainty corresponds to 3\,$\sigma$ to provide a broader range of realistic values for the total mass.  We also placed an arbitrary upper limit of $P <$ 5000 yr on the orbital period.  The best fit and range of orbital parameters that represent the motion of the center of mass of T Tau S relative to T Tau N are listed in Table~\ref{tab.wide} and plotted in Figure~\ref{fig.wide}.  These are consistent with the range of orbits for the wide pair found by \citet{kohler16}.  

If the effective temperature of T Tau N is lower than the value implied by the optical spectral type, then this would lead to a smaller mass for T Tau N.  However, changing the total mass constraint based on where the effective temperature implied by the infrared spectral type of T Tau N intersects the evolutionary tracks in Figure~\ref{fig.hrd} produces a similar range of possible orbital parameters for the wide N,S orbit.  Moreover, the resulting masses of T Tau Sa and Sb change by only 0.003 $M_\odot$, well within the 1$\sigma$ uncertainty intervals reported in Table~\ref{tab.mass}.

\section{Variability in the T Tau System} 

\begin{figure}
  \begin{center}
	\scalebox{0.48}{\includegraphics{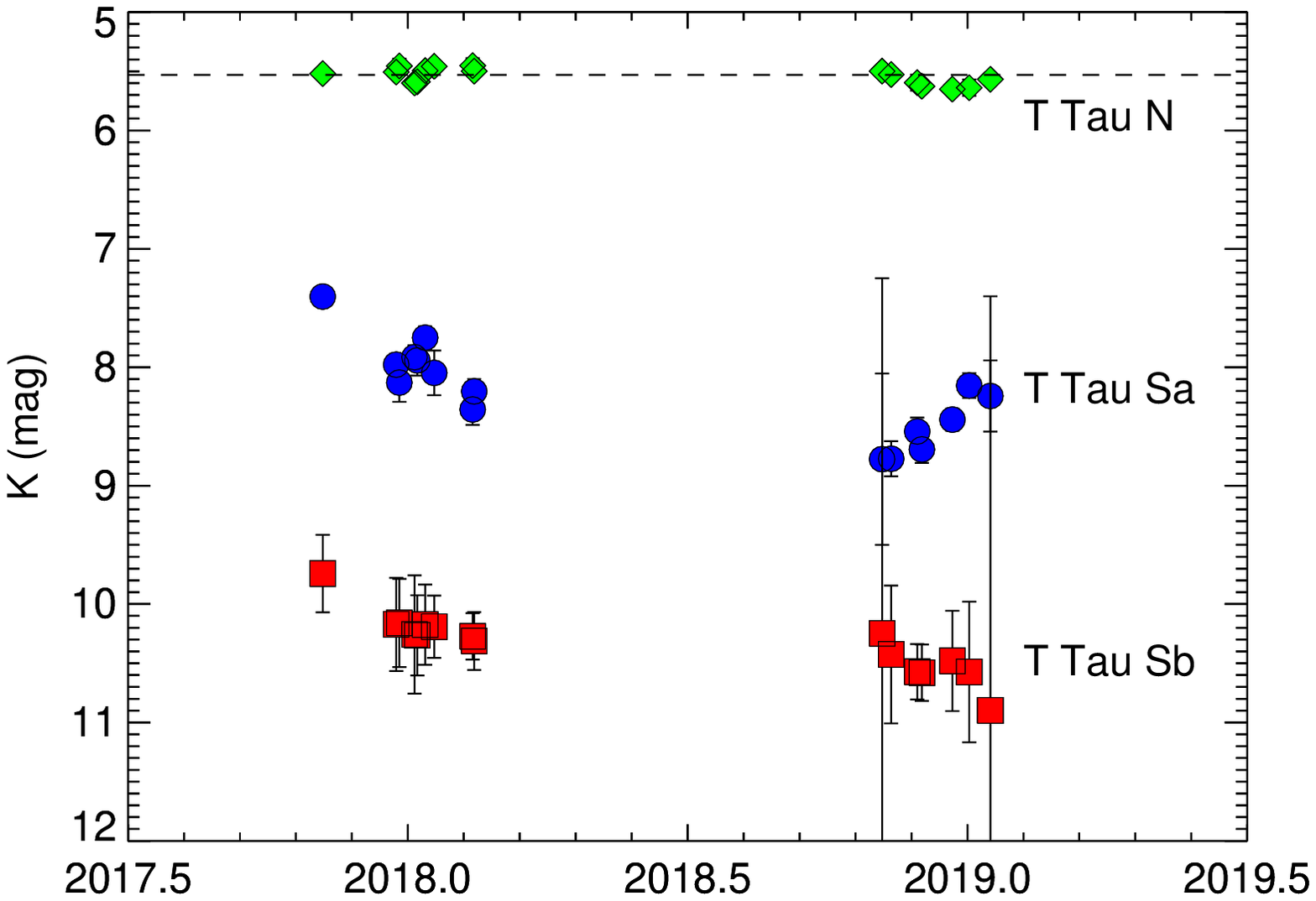}} \\
	\scalebox{0.48}{\includegraphics{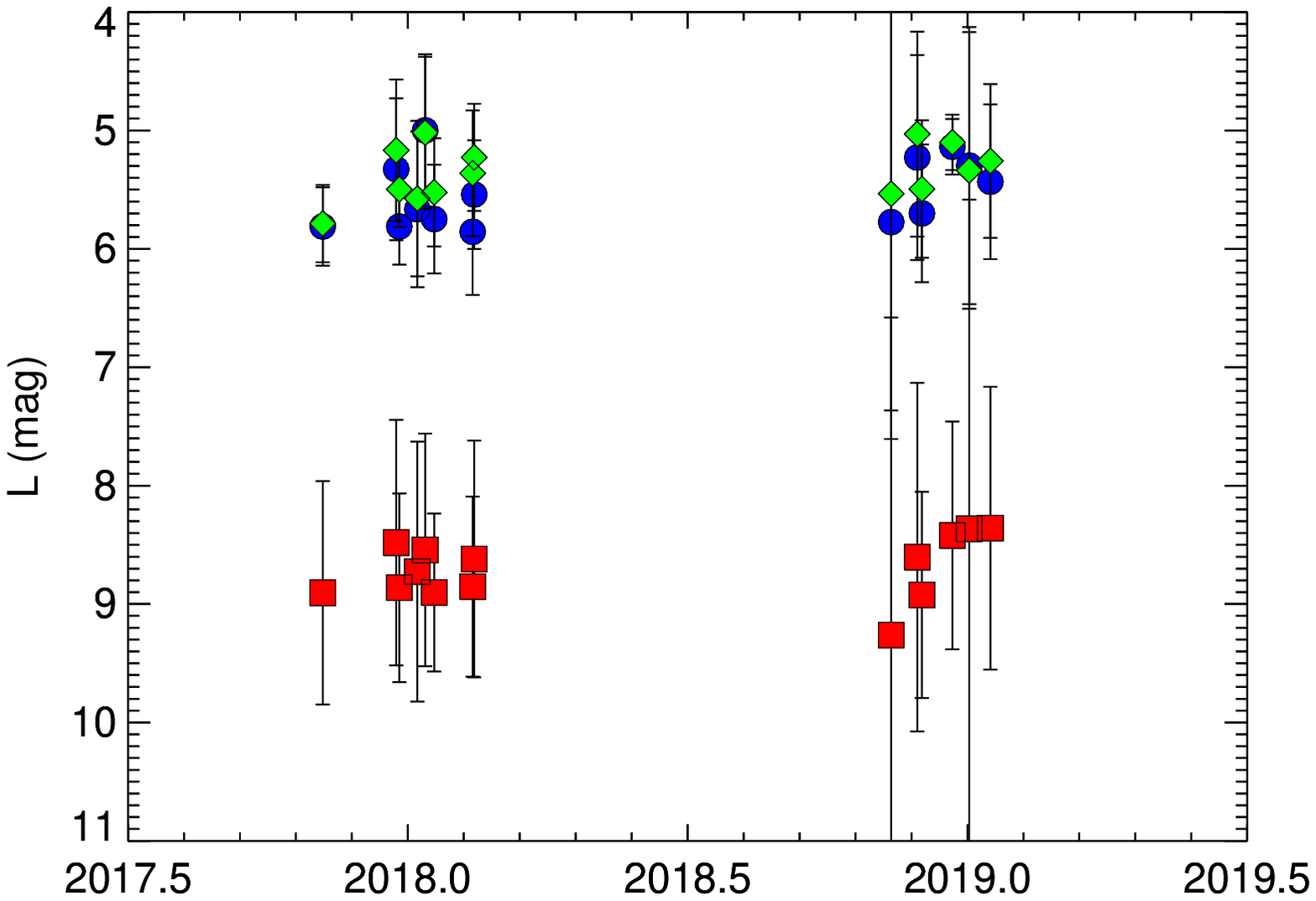}} \\
	\scalebox{0.48}{\includegraphics{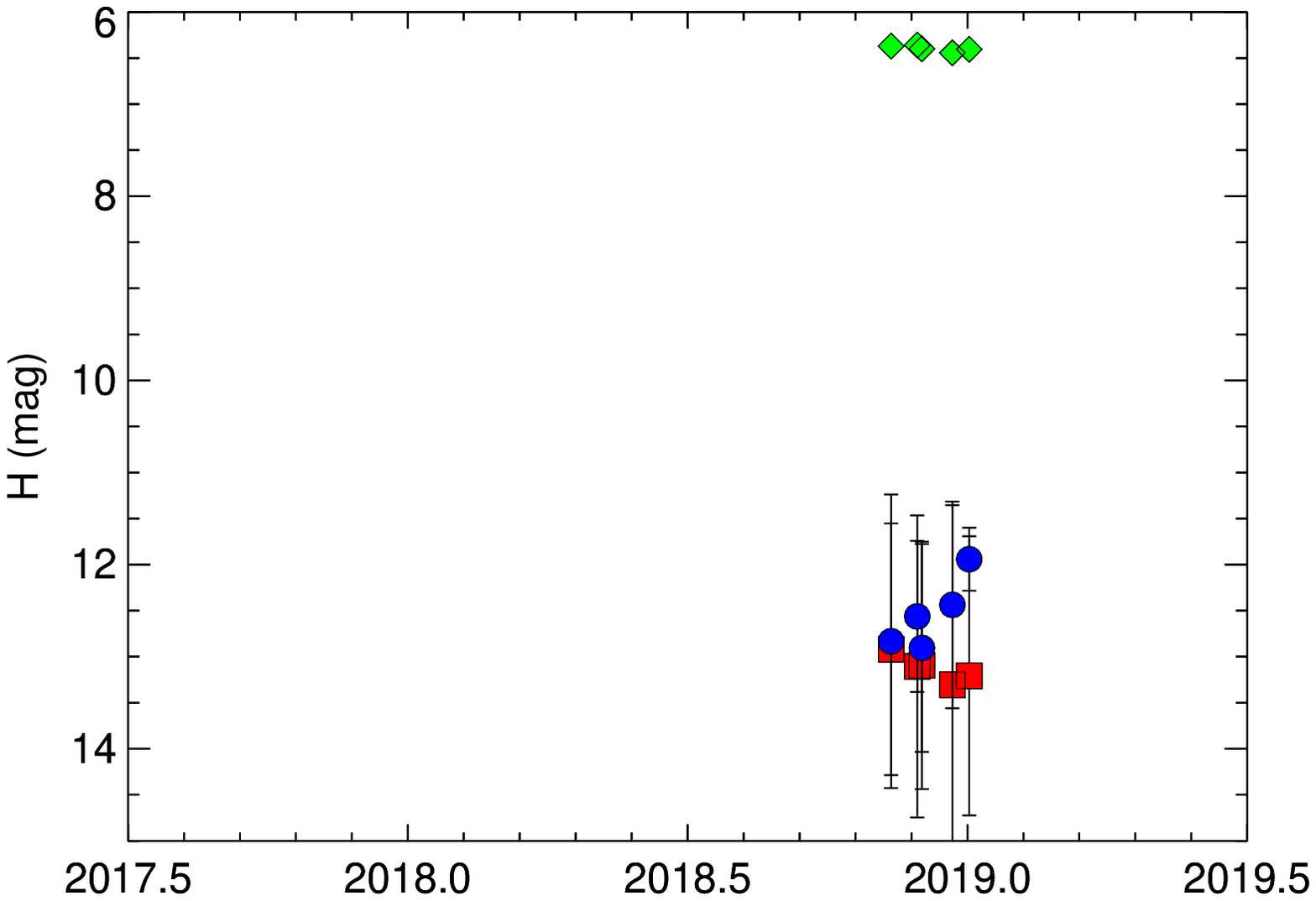}}
        \caption{Magnitudes of T Tau N (green diamonds), Sa (blue circles), and Sb (red squares) based on Gemini AO observations on nights when the near-infrared flux standard HD 22686 was also observed (Sect.~\ref{sect.phot}).  The Kcon, Hcon, and Br$\alpha$ filters were used as a proxy for the $KHL$ fluxes (top, middle, and bottom panels, respectively).  The dashed line in the top panel corresponds to the mean magnitude of K = 5.53 $\pm$ 0.03 determined by \citet{beck04}. \label{fig.phot}}
  \end{center}
\end{figure}

The absolute photometric measurements from the Gemini observations (Table~\ref{tab.phot}) are plotted in Figure~\ref{fig.phot}.  The $K$-band magnitude of T Tau N remained steady with a range of 5.45$-$5.65 mag and an average value of $K$ = 5.54 $\pm$ 0.07 mag.  This is consistent with the results reported by \citet{beck04} who found that the infrared flux of T Tau N remained constant from 1994 to 2002, with an average magnitude of $K = 5.53 \pm 0.03$ mag.  The uncertainties at $H$ and $L$ are larger than at $K$ because of the small flux of Sa and Sb in the H-band ($\gtrsim$ 6 mag fainter than T Tau N), and the lower angular resolution in the $L$-band. T Tau N and Sa are similar in brightness in the $L$-band, but Sb is much fainter.

\begin{figure}
\plotone{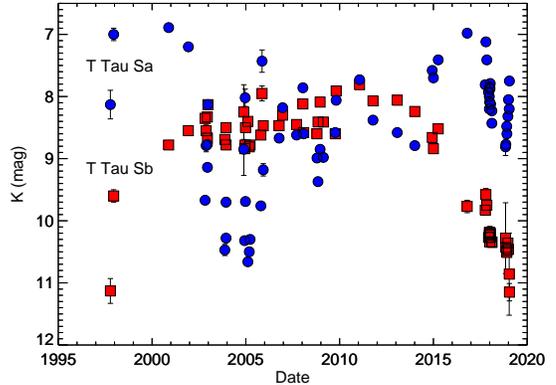}
\caption{Variability of T Tau Sa (blue circles) and Sb (red squares) relative to T Tau N in the $K$-band.  We derived the component magnitudes from the flux ratios and assumed a constant magnitude of $K = 5.53 \pm 0.03$ mag for T Tau N \citep{beck04}.  The flux ratios are from Table~\ref{tab.sepPA} and measurements in the literature \citep{koresko00,duchene02,duchene05,duchene06,furlan03,beck04,mayama06,schaefer06,schaefer14,vanboekel10,kasper16}.
\label{fig.mag}}
\end{figure}

We can expand the time-frame of the variability measurements by assuming an average magnitude of $K = 5.53 \pm 0.03$ mag for T Tau N \citep{beck04} and converting the relative flux ratios between the three components into magnitudes.  The long-term brightness variations of T Tau Sa and Sb in the $K$-band are plotted in Figure~\ref{fig.mag}.  From 2015 to 2019, T Tau Sa experienced a dramatic increase in brightness, becoming $\sim$ 2 mag brighter than Sb, continuing the brightening trend reported initially by \citet{csepany15} and \citet{kasper16}.  According to the Gemini observations that were taken with higher temporal sampling, the $K$-band magnitude of T Tau Sa dropped in early 2018 and then began rising again in late 2018 through early 2019.  The $K$-band magnitude of T Tau Sb steadily decreased by $\sim$ 2.6 mag over the 2015$-$2019 interval. 

\section{Conclusions}

Based on our recent AO imaging of the T Tau triple system, combined with prior measurements in the literature, we fit the orbital motion of T Tau Sb relative to Sa and modeled the astrometric motion of their center of mass relative to T Tau N.  Using the distance of 143.74 $\pm$ 1.22 pc \citep{bailerjones18}, we derived dynamical masses of $M_{\rm Sa} = 2.05 \pm 0.14$ M$_\odot$ and $M_{\rm Sb} = 0.43 \pm 0.06$ M$_\odot$.  The orbital parameters, mass ratio, and masses are consistent within their uncertainties with the values computed by \citet{kohler16}.  However, the current uncertainties in the orbital parameters are significantly smaller thanks to the improved orbital coverage obtained over the past four years.  

The fluxes derived from the AO images show that the $K$-band flux of T Tau N has remained steady between late 2017 and early 2019, with an average value of $K$ = 5.54 $\pm$ 0.07 mag.  T Tau Sa is again brighter than Sb, but its $K$-band brightness varied dramatically in the past four years between 7.0 to 8.8 mag over timescales of a few months.  On the other hand, T Tau Sb faded steadily from $K$ = 8.5 to 11.1 mag over four years.  In a forthcoming paper, T.~Beck et al.\ (in prep) investigate the link between the variability, orbital motion, circumstellar emission, and outflows in the system.

\acknowledgements

We thank the staff at the Keck and Gemini observatories for their support during the observations. We also thank the referee for providing feedback that improved the paper.  G.H.S.\ and L.P. acknowledge support from NASA Keck PI Data Awards administered by the NASA Exoplanet Science Institute.  Additional support was provided through the National Science Foundation (AST-1636624 for G.H.S. and AST-1518081 for L.P.).  Some of the data presented herein were obtained at the W. M. Keck Observatory from telescope time allocated to the National Aeronautics and Space Administration through the agency's scientific partnership with the California Institute of Technology and the University of California. The Observatory was made possible by the generous financial support of the W. M. Keck Foundation.  Some of the observations were obtained at the Gemini Observatory (GN-2017B-Q-29, GN-2018B-Q-137), which is operated by the Association of Universities for Research in Astronomy, Inc., under a cooperative agreement with the NSF on behalf of the Gemini partnership: the National Science Foundation (United States), National Research Council (Canada), CONICYT (Chile), Ministerio de Ciencia, Tecnolog\'{i}a e Innovaci\'{o}n Productiva (Argentina), Minist\'{e}rio da Ci\^{e}ncia, Tecnologia e Inova\c{c}\~{a}o (Brazil), and Korea Astronomy and Space Science Institute (Republic of Korea).  Time at Gemini was granted through the time allocation process at the National Optical Astronomical Observatory (NOAO Prop.~ID: 2017B-0280, 2018B-0321; PI: Schaefer).  The data were downloaded through the Gemini Observatory Archive.  We wish to recognize and acknowledge the significant cultural role that the summit of Maunakea has within the indigenous Hawaiian community.  We are sincerely grateful for the opportunity to conduct these observations from the mountain. This research has made use of the SIMBAD database and the VizieR catalog access tool, CDS, Strasbourg, France.

\facilities{Keck:II (NIRC2), Gemini:Gillett (NIRI)} 

\bibliography{ms}

\begin{thebibliography}{}
\expandafter\ifx\csname natexlab\endcsname\relax\def\natexlab#1{#1}\fi

\bibitem[{{Bailer-Jones} {et~al.}(2018){Bailer-Jones}, {Rybizki}, {Fouesneau},
  {Mantelet}, \& {Andrae}}]{bailerjones18}
{Bailer-Jones}, C.~A.~L., {Rybizki}, J., {Fouesneau}, M., {Mantelet}, G., \&
  {Andrae}, R. 2018, \aj, 156, 58

\bibitem[{{Beck} {et~al.}(2004){Beck}, {Schaefer}, {Simon}, {Prato}, {Stoesz},
  \& {Howell}}]{beck04}
{Beck}, T.~L., {Schaefer}, G.~H., {Simon}, M., {et~al.} 2004, \apj, 614, 235

\bibitem[{{Cs{\'e}p{\'a}ny} {et~al.}(2015){Cs{\'e}p{\'a}ny}, {van den Ancker},
  {{\'A}brah{\'a}m}, {Brandner}, \& {Hormuth}}]{csepany15}
{Cs{\'e}p{\'a}ny}, G., {van den Ancker}, M., {{\'A}brah{\'a}m}, P., {Brandner},
  W., \& {Hormuth}, F. 2015, \aap, 578, L9

\bibitem[{{Duch{\^e}ne} {et~al.}(2006){Duch{\^e}ne}, {Beust}, {Adjali},
  {Konopacky}, \& {Ghez}}]{duchene06}
{Duch{\^e}ne}, G., {Beust}, H., {Adjali}, F., {Konopacky}, Q.~M., \& {Ghez},
  A.~M. 2006, \aap, 457, L9

\bibitem[{{Duch{\^e}ne} {et~al.}(2002){Duch{\^e}ne}, {Ghez}, \&
  {McCabe}}]{duchene02}
{Duch{\^e}ne}, G., {Ghez}, A.~M., \& {McCabe}, C. 2002, \apj, 568, 771

\bibitem[{{Duch{\^e}ne} {et~al.}(2005){Duch{\^e}ne}, {Ghez}, {McCabe}, \&
  {Ceccarelli}}]{duchene05}
{Duch{\^e}ne}, G., {Ghez}, A.~M., {McCabe}, C., \& {Ceccarelli}, C. 2005, \apj,
  628, 832

\bibitem[{{Dyck} {et~al.}(1982){Dyck}, {Simon}, \& {Zuckerman}}]{dyck82}
{Dyck}, H.~M., {Simon}, T., \& {Zuckerman}, B. 1982, \apjl, 255, L103

\bibitem[{{Feiden}(2016)}]{feiden16}
{Feiden}, G.~A. 2016, \aap, 593, A99

\bibitem[{{Furlan} {et~al.}(2003){Furlan}, {Forrest}, {Watson}, {Uchida},
  {Brandl}, {Keller}, \& {Herter}}]{furlan03}
{Furlan}, E., {Forrest}, W.~J., {Watson}, D.~M., {et~al.} 2003, \apjl, 596, L87

\bibitem[{{Gaia Collaboration} {et~al.}(2018){Gaia Collaboration}, {Brown},
  {Vallenari}, {Prusti}, {de Bruijne}, {Babusiaux}, {Bailer-Jones}, {Biermann},
  {Evans}, {Eyer}, {Jansen}, {Jordi}, {Klioner}, {Lammers}, {Lindegren},
  {Luri}, {Mignard}, {Panem}, {Pourbaix}, {Randich}, {Sartoretti}, {Siddiqui},
  {Soubiran}, {van Leeuwen}, {Walton}, {Arenou}, {Bastian}, {Cropper},
  {Drimmel}, {Katz}, {Lattanzi}, {Bakker}, {Cacciari}, {Casta{\~n}eda},
  {Chaoul}, {Cheek}, {De Angeli}, {Fabricius}, {Guerra}, {Holl}, {Masana},
  {Messineo}, {Mowlavi}, {Nienartowicz}, {Panuzzo}, {Portell}, {Riello},
  {Seabroke}, {Tanga}, {Th{\'e}venin}, {Gracia-Abril}, {Comoretto},
  {Garcia-Reinaldos}, {Teyssier}, {Altmann}, {Andrae}, {Audard},
  {Bellas-Velidis}, {Benson}, {Berthier}, {Blomme}, {Burgess}, {Busso},
  {Carry}, {Cellino}, {Clementini}, {Clotet}, {Creevey}, {Davidson}, {De
  Ridder}, {Delchambre}, {Dell'Oro}, {Ducourant},
  {Fern{\'a}ndez-Hern{\'a}ndez}, {Fouesneau}, {Fr{\'e}mat}, {Galluccio},
  {Garc{\'\i}a-Torres}, {Gonz{\'a}lez-N{\'u}{\~n}ez}, {Gonz{\'a}lez-Vidal},
  {Gosset}, {Guy}, {Halbwachs}, {Hambly}, {Harrison}, {Hern{\'a}ndez},
  {Hestroffer}, {Hodgkin}, {Hutton}, {Jasniewicz}, {Jean-Antoine-Piccolo},
  {Jordan}, {Korn}, {Krone-Martins}, {Lanzafame}, {Lebzelter}, {L{\"o}ffler},
  {Manteiga}, {Marrese}, {Mart{\'\i}n-Fleitas}, {Moitinho}, {Mora}, {Muinonen},
  {Osinde}, {Pancino}, {Pauwels}, {Petit}, {Recio-Blanco}, {Richards},
  {Rimoldini}, {Robin}, {Sarro}, {Siopis}, {Smith}, {Sozzetti}, {S{\"u}veges},
  {Torra}, {van Reeven}, {Abbas}, {Abreu Aramburu}, {Accart}, {Aerts},
  {Altavilla}, {{\'A}lvarez}, {Alvarez}, {Alves}, {Anderson}, {Andrei},
  {Anglada Varela}, {Antiche}, {Antoja}, {Arcay}, {Astraatmadja}, {Bach},
  {Baker}, {Balaguer-N{\'u}{\~n}ez}, {Balm}, {Barache}, {Barata}, {Barbato},
  {Barblan}, {Barklem}, {Barrado}, {Barros}, {Barstow}, {Bartholom{\'e}
  Mu{\~n}oz}, {Bassilana}, {Becciani}, {Bellazzini}, {Berihuete}, {Bertone},
  {Bianchi}, {Bienaym{\'e}}, {Blanco-Cuaresma}, {Boch}, {Boeche}, {Bombrun},
  {Borrachero}, {Bossini}, {Bouquillon}, {Bourda}, {Bragaglia}, {Bramante},
  {Breddels}, {Bressan}, {Brouillet}, {Br{\"u}semeister}, {Brugaletta},
  {Bucciarelli}, {Burlacu}, {Busonero}, {Butkevich}, {Buzzi}, {Caffau},
  {Cancelliere}, {Cannizzaro}, {Cantat-Gaudin}, {Carballo}, {Carlucci},
  {Carrasco}, {Casamiquela}, {Castellani}, {Castro-Ginard}, {Charlot},
  {Chemin}, {Chiavassa}, {Cocozza}, {Costigan}, {Cowell}, {Crifo}, {Crosta},
  {Crowley}, {Cuypers}, {Dafonte}, {Damerdji}, {Dapergolas}, {David}, {David},
  {de Laverny}, {De Luise}, {De March}, {de Martino}, {de Souza}, {de Torres},
  {Debosscher}, {del Pozo}, {Delbo}, {Delgado}, {Delgado}, {Di Matteo},
  {Diakite}, {Diener}, {Distefano}, {Dolding}, {Drazinos}, {Dur{\'a}n},
  {Edvardsson}, {Enke}, {Eriksson}, {Esquej}, {Eynard Bontemps}, {Fabre},
  {Fabrizio}, {Faigler}, {Falc{\~a}o}, {Farr{\`a}s Casas}, {Federici},
  {Fedorets}, {Fernique}, {Figueras}, {Filippi}, {Findeisen}, {Fonti},
  {Fraile}, {Fraser}, {Fr{\'e}zouls}, {Gai}, {Galleti}, {Garabato},
  {Garc{\'\i}a-Sedano}, {Garofalo}, {Garralda}, {Gavel}, {Gavras}, {Gerssen},
  {Geyer}, {Giacobbe}, {Gilmore}, {Girona}, {Giuffrida}, {Glass}, {Gomes},
  {Granvik}, {Gueguen}, {Guerrier}, {Guiraud}, {Guti{\'e}rrez-S{\'a}nchez},
  {Haigron}, {Hatzidimitriou}, {Hauser}, {Haywood}, {Heiter}, {Helmi}, {Heu},
  {Hilger}, {Hobbs}, {Hofmann}, {Holland}, {Huckle}, {Hypki}, {Icardi},
  {Jan{\ss}en}, {Jevardat de Fombelle}, {Jonker}, {Juh{\'a}sz}, {Julbe},
  {Karampelas}, {Kewley}, {Klar}, {Kochoska}, {Kohley}, {Kolenberg},
  {Kontizas}, {Kontizas}, {Koposov}, {Kordopatis}, {Kostrzewa-Rutkowska},
  {Koubsky}, {Lambert}, {Lanza}, {Lasne}, {Lavigne}, {Le Fustec}, {Le
  Poncin-Lafitte}, {Lebreton}, {Leccia}, {Leclerc}, {Lecoeur-Taibi},
  {Lenhardt}, {Leroux}, {Liao}, {Licata}, {Lindstr{\o}m}, {Lister}, {Livanou},
  {Lobel}, {L{\'o}pez}, {Managau}, {Mann}, {Mantelet}, {Marchal}, {Marchant},
  {Marconi}, {Marinoni}, {Marschalk{\'o}}, {Marshall}, {Martino}, {Marton},
  {Mary}, {Massari}, {Matijevi{\v{c}}}, {Mazeh}, {McMillan}, {Messina},
  {Michalik}, {Millar}, {Molina}, {Molinaro}, {Moln{\'a}r}, {Montegriffo},
  {Mor}, {Morbidelli}, {Morel}, {Morris}, {Mulone}, {Muraveva}, {Musella},
  {Nelemans}, {Nicastro}, {Noval}, {O'Mullane}, {Ord{\'e}novic},
  {Ord{\'o}{\~n}ez-Blanco}, {Osborne}, {Pagani}, {Pagano}, {Pailler},
  {Palacin}, {Palaversa}, {Panahi}, {Pawlak}, {Piersimoni}, {Pineau}, {Plachy},
  {Plum}, {Poggio}, {Poujoulet}, {Pr{\v{s}}a}, {Pulone}, {Racero}, {Ragaini},
  {Rambaux}, {Ramos-Lerate}, {Regibo}, {Reyl{\'e}}, {Riclet}, {Ripepi}, {Riva},
  {Rivard}, {Rixon}, {Roegiers}, {Roelens}, {Romero-G{\'o}mez}, {Rowell},
  {Royer}, {Ruiz-Dern}, {Sadowski}, {Sagrist{\`a} Sell{\'e}s}, {Sahlmann},
  {Salgado}, {Salguero}, {Sanna}, {Santana-Ros}, {Sarasso}, {Savietto},
  {Schultheis}, {Sciacca}, {Segol}, {Segovia}, {S{\'e}gransan}, {Shih},
  {Siltala}, {Silva}, {Smart}, {Smith}, {Solano}, {Solitro}, {Sordo}, {Soria
  Nieto}, {Souchay}, {Spagna}, {Spoto}, {Stampa}, {Steele},
  {Steidelm{\"u}ller}, {Stephenson}, {Stoev}, {Suess}, {Surdej}, {Szabados},
  {Szegedi-Elek}, {Tapiador}, {Taris}, {Tauran}, {Taylor}, {Teixeira},
  {Terrett}, {Teyssand ier}, {Thuillot}, {Titarenko}, {Torra Clotet}, {Turon},
  {Ulla}, {Utrilla}, {Uzzi}, {Vaillant}, {Valentini}, {Valette}, {van Elteren},
  {Van Hemelryck}, {van Leeuwen}, {Vaschetto}, {Vecchiato}, {Veljanoski},
  {Viala}, {Vicente}, {Vogt}, {von Essen}, {Voss}, {Votruba}, {Voutsinas},
  {Walmsley}, {Weiler}, {Wertz}, {Wevers}, {Wyrzykowski}, {Yoldas},
  {{\v{Z}}erjal}, {Ziaeepour}, {Zorec}, {Zschocke}, {Zucker}, {Zurbach}, \&
  {Zwitter}}]{gaia18}
{Gaia Collaboration}, {Brown}, A.~G.~A., {Vallenari}, A., {et~al.} 2018, \aap,
  616, A1

\bibitem[{{Galli} {et~al.}(2018){Galli}, {Loinard}, {Ortiz-L{\'e}on},
  {Kounkel}, {Dzib}, {Mioduszewski}, {Rodr{\'\i}guez}, {Hartmann}, {Teixeira},
  {Torres}, {Rivera}, {Boden}, {Evans}, {Brice{\~n}o}, {Tobin}, \&
  {Heyer}}]{galli18}
{Galli}, P. A.~B., {Loinard}, L., {Ortiz-L{\'e}on}, G.~N., {et~al.} 2018, \apj,
  859, 33

\bibitem[{{Galli} {et~al.}(2019){Galli}, {Loinard}, {Bouy}, {Sarro},
  {Ortiz-Le{\'o}n}, {Dzib}, {Olivares}, {Heyer}, {Hernandez},
  {Rom{\'a}n-Z{\'u}{\~n}iga}, {Kounkel}, \& {Covey}}]{galli19}
{Galli}, P.~A.~B., {Loinard}, L., {Bouy}, H., {et~al.} 2019, \aap, 630, A137

\bibitem[{{Guetter} {et~al.}(2003){Guetter}, {Vrba}, {Henden}, \&
  {Luginbuhl}}]{guetter03}
{Guetter}, H.~H., {Vrba}, F.~J., {Henden}, A.~A., \& {Luginbuhl}, C.~B. 2003,
  \aj, 125, 3344

\bibitem[{{Gully-Santiago} {et~al.}(2017){Gully-Santiago}, {Herczeg},
  {Czekala}, {Somers}, {Grankin}, {Covey}, {Donati}, {Alencar}, {Hussain},
  {Shappee}, {Mace}, {Lee}, {Holoien}, {Jose}, \& {Liu}}]{gully-santiago17}
{Gully-Santiago}, M.~A., {Herczeg}, G.~J., {Czekala}, I., {et~al.} 2017, \apj,
  836, 200

\bibitem[{{Hodapp} {et~al.}(2003){Hodapp}, {Jensen}, {Irwin}, {Yamada},
  {Chung}, {Fletcher}, {Robertson}, {Hora}, {Simons}, {Mays}, {Nolan}, {Bec},
  {Merrill}, \& {Fowler}}]{hodapp03}
{Hodapp}, K.~W., {Jensen}, J.~B., {Irwin}, E.~M., {et~al.} 2003, \pasp, 115,
  1388

\bibitem[{{Joy}(1945)}]{joy45}
{Joy}, A.~H. 1945, \apj, 102, 168

\bibitem[{{Kasper} {et~al.}(2016){Kasper}, {Santhakumari}, {Herbst}, \&
  {K{\"o}hler}}]{kasper16}
{Kasper}, M., {Santhakumari}, K.~K.~R., {Herbst}, T.~M., \& {K{\"o}hler}, R.
  2016, \aap, 593, A50

\bibitem[{{K{\"o}hler} {et~al.}(2000){K{\"o}hler}, {Kasper}, \&
  {Herbst}}]{kohler00}
{K{\"o}hler}, R., {Kasper}, M., \& {Herbst}, T. 2000, in Birth and Evolution of
  Binary Stars: Poster Proceedings of IAU Symposium No. 200 on the Formation of
  Binary Stars, ed. B.~{Reipurth} \& H.~{Zinnecker} (Potsdam, Germany:
  Astrophysikalisches Institut Potsdam), 63

\bibitem[{{K{\"o}hler} {et~al.}(2016){K{\"o}hler}, {Kasper}, {Herbst},
  {Ratzka}, \& {Bertrang}}]{kohler16}
{K{\"o}hler}, R., {Kasper}, M., {Herbst}, T.~M., {Ratzka}, T., \& {Bertrang},
  G.~H.~M. 2016, \aap, 587, A35

\bibitem[{{K{\"o}hler} {et~al.}(2008){K{\"o}hler}, {Ratzka}, {Herbst}, \&
  {Kasper}}]{kohler08}
{K{\"o}hler}, R., {Ratzka}, T., {Herbst}, T.~M., \& {Kasper}, M. 2008, \aap,
  482, 929

\bibitem[{{Koresko}(2000)}]{koresko00}
{Koresko}, C.~D. 2000, \apjl, 531, L147

\bibitem[{{Koresko} {et~al.}(1997){Koresko}, {Herbst}, \&
  {Leinert}}]{koresko97}
{Koresko}, C.~D., {Herbst}, T.~M., \& {Leinert}, C. 1997, \apj, 480, 741

\bibitem[{{Leggett} {et~al.}(2003){Leggett}, {Hawarden}, {Currie}, {Adamson},
  {Carroll}, {Kerr}, {Kuhn}, {Seigar}, {Varricatt}, \& {Wold}}]{leggett03}
{Leggett}, S.~K., {Hawarden}, T.~G., {Currie}, M.~J., {et~al.} 2003, \mnras,
  345, 144

\bibitem[{{Loinard} {et~al.}(2007){Loinard}, {Torres}, {Mioduszewski},
  {Rodr{\'{\i}}guez}, {Gonz{\'a}lez-L{\'o}pezlira}, {Lachaume}, {V{\'a}zquez},
  \& {Gonz{\'a}lez}}]{loinard07}
{Loinard}, L., {Torres}, R.~M., {Mioduszewski}, A.~J., {et~al.} 2007, \apj,
  671, 546

\bibitem[{{Luhman}(2018)}]{luhman18}
{Luhman}, K.~L. 2018, \aj, 156, 271

\bibitem[{{Manara} {et~al.}(2019){Manara}, {Tazzari}, {Long}, {Herczeg},
  {Lodato}, {Rota}, {Cazzoletti}, {van der Plas}, {Pinilla}, {Dipierro},
  {Edwards}, {Harsono}, {Johnstone}, {Liu}, {Menard}, {Nisini}, {Ragusa},
  {Boehler}, \& {Cabrit}}]{manara19}
{Manara}, C.~F., {Tazzari}, M., {Long}, F., {et~al.} 2019, \aap, 628, A95

\bibitem[{{Mayama} {et~al.}(2006){Mayama}, {Tamura}, {Hayashi}, {Itoh},
  {Fukagawa}, {Suto}, {Ishii}, {Murakawa}, {Oasa}, {Hayashi}, {Yamashita},
  {Morino}, {Oya}, {Naoi}, {Pyo}, {Nishikawa}, {Kudo}, {Usuda}, {Ando},
  {Miyama}, \& {Kaifu}}]{mayama06}
{Mayama}, S., {Tamura}, M., {Hayashi}, M., {et~al.} 2006, \pasj, 58, 375

\bibitem[{{Pecaut} \& {Mamajek}(2013)}]{pecaut13}
{Pecaut}, M.~J., \& {Mamajek}, E.~E. 2013, \apjs, 208, 9

\bibitem[{{Ratzka} {et~al.}(2009){Ratzka}, {Schegerer}, {Leinert},
  {{\'A}brah{\'a}m}, {Henning}, {Herbst}, {K{\"o}hler}, {Wolf}, \&
  {Zinnecker}}]{ratzka09}
{Ratzka}, T., {Schegerer}, A.~A., {Leinert}, C., {et~al.} 2009, \aap, 502, 623

\bibitem[{{Schaefer} {et~al.}(2014){Schaefer}, {Prato}, {Simon}, \&
  {Patience}}]{schaefer14}
{Schaefer}, G.~H., {Prato}, L., {Simon}, M., \& {Patience}, J. 2014, \aj, 147,
  157

\bibitem[{{Schaefer} {et~al.}(2012){Schaefer}, {Prato}, {Simon}, \&
  {Zavala}}]{schaefer12}
{Schaefer}, G.~H., {Prato}, L., {Simon}, M., \& {Zavala}, R.~T. 2012, \apj,
  756, 120

\bibitem[{{Schaefer} {et~al.}(2006){Schaefer}, {Simon}, {Beck}, {Nelan}, \&
  {Prato}}]{schaefer06}
{Schaefer}, G.~H., {Simon}, M., {Beck}, T.~L., {Nelan}, E., \& {Prato}, L.
  2006, \aj, 132, 2618

\bibitem[{{Service} {et~al.}(2016){Service}, {Lu}, {Campbell}, {Sitarski},
  {Ghez}, \& {Anderson}}]{service16}
{Service}, M., {Lu}, J.~R., {Campbell}, R., {et~al.} 2016, \pasp, 128, 095004

\bibitem[{{Skemer} {et~al.}(2008){Skemer}, {Close}, {Hinz}, {Hoffmann},
  {Kenworthy}, \& {Miller}}]{skemer08}
{Skemer}, A.~J., {Close}, L.~M., {Hinz}, P.~M., {et~al.} 2008, \apj, 676, 1082

\bibitem[{{Stapelfeldt} {et~al.}(1998){Stapelfeldt}, {Burrows}, {Krist},
  {Watson}, {Ballester}, {Clarke}, {Crisp}, {Evans}, {Gallagher}, {Griffiths},
  {Hester}, {Hoessel}, {Holtzman}, {Mould}, {Scowen}, {Trauger}, \&
  {Westphal}}]{stapelfeldt98}
{Stapelfeldt}, K.~R., {Burrows}, C.~J., {Krist}, J.~E., {et~al.} 1998, \apj,
  508, 736

\bibitem[{{Tokunaga} {et~al.}(2002){Tokunaga}, {Simons}, \&
  {Vacca}}]{tokunaga02}
{Tokunaga}, A.~T., {Simons}, D.~A., \& {Vacca}, W.~D. 2002, \pasp, 114, 180

\bibitem[{{van Boekel} {et~al.}(2010){van Boekel}, {Juh{\'a}sz}, {Henning},
  {K{\"o}hler}, {Ratzka}, {Herbst}, {Bouwman}, \& {Kley}}]{vanboekel10}
{van Boekel}, R., {Juh{\'a}sz}, A., {Henning}, T., {et~al.} 2010, \aap, 517,
  A16

\bibitem[{{Wizinowich} {et~al.}(2000){Wizinowich}, {Acton}, {Shelton},
  {Stomski}, {Gathright}, {Ho}, {Lupton}, {Tsubota}, {Lai}, {Max}, {Brase},
  {An}, {Avicola}, {Olivier}, {Gavel}, {Macintosh}, {Ghez}, \&
  {Larkin}}]{wizinowich00new}
{Wizinowich}, P., {Acton}, D.~S., {Shelton}, C., {et~al.} 2000, \pasp, 112, 315

\bibitem[{{Yelda} {et~al.}(2010){Yelda}, {Lu}, {Ghez}, {Clarkson}, {Anderson},
  {Do}, \& {Matthews}}]{yelda10}
{Yelda}, S., {Lu}, J.~R., {Ghez}, A.~M., {et~al.} 2010, \apj, 725, 331

\end{thebibliography}

\mbox

\clearpage

\onecolumngrid
\begin{deluxetable}{llllllll} 
\tabletypesize{\scriptsize}
\tablewidth{0pt}
\tablecaption{Near-IR Adaptive Optics Measurements of T Tau Triple System \label{tab.sepPA}} 
\tablehead{
\colhead{UT Date} & \colhead{JY} & \colhead{Pair} & \colhead{$\rho$(mas)} & \colhead{P.A.($\degr$)} & \colhead{Filter} & \colhead{Flux Ratio} &\colhead{Tel\tablenotemark{a}}}
\startdata 
2015Jan01 & 2015.0000  & Sa,Sb &   110.34  $\pm$   0.55  &   345.52  $\pm$   0.29  & Kcont & 0.3491$\pm$0.0078 & K \\
          &            &       &                         &                         & Hcont & 2.9314$\pm$0.5648 & K \\
2015Apr05 & 2015.2573  & Sa,Sb &   108.09  $\pm$   0.40  &   346.94  $\pm$   0.21  & Kcont & 0.3612$\pm$0.0067 & K \\
2016Oct20 & 2016.8021  & Sa,Sb &    96.79  $\pm$   2.09  &   357.88  $\pm$   1.24  & Kcont & 0.0768$\pm$0.0066 & K \\
          &            &       &                         &                         & Hcont & 0.1572$\pm$0.0372 & K \\
2017Oct05 & 2017.7605  & Sa,Sb &    91.89  $\pm$   6.76  &     6.94  $\pm$   4.21  &  Kcon & 0.1554$\pm$0.0127 & G \\
          &            &       &                         &                         &   BrA & 0.0823$\pm$0.0189 & G \\
2017Oct19 & 2017.7989  & Sa,Sb &    91.60  $\pm$   3.24  &     6.25  $\pm$   2.03  &  Kcon & 0.1033$\pm$0.0099 & G \\
          &            &       &                         &                         &   BrA & 0.0527$\pm$0.0081 & G \\
2017Nov06 & 2017.8479  & Sa,Sb &    92.00  $\pm$   2.99  &     6.91  $\pm$   1.86  &  Kcon & 0.1160$\pm$0.0073 & G \\
          &            &       &                         &                         &   BrA & 0.0579$\pm$0.0091 & G \\
2017Dec09 & 2017.9382  & Sa,Sb &    90.24  $\pm$   2.67  &     7.50  $\pm$   1.69  &  Kcon & 0.1152$\pm$0.0083 & G \\
          &            &       &                         &                         &   BrA & 0.0492$\pm$0.0083 & G \\
2017Dec24 & 2017.9792  & Sa,Sb &    89.40  $\pm$   4.35  &     8.74  $\pm$   2.79  &  Kcon & 0.1326$\pm$0.0094 & G \\
          &            &       &                         &                         &   BrA & 0.0548$\pm$0.0082 & G \\
2017Dec26 & 2017.9848  & Sa,Sb &    89.36  $\pm$   2.72  &     8.87  $\pm$   1.74  &  Kcon & 0.1543$\pm$0.0115 & G \\
          &            &       &                         &                         &   BrA & 0.0602$\pm$0.0079 & G \\
2017Dec31 & 2017.9983  & Sa,Sb &    87.34  $\pm$   2.81  &     8.68  $\pm$   1.84  &  Kcon & 0.1260$\pm$0.0097 & G \\
2018Jan05 & 2018.0120  & Sa,Sb &    87.74  $\pm$   2.59  &     8.38  $\pm$   1.69  &  Kcon & 0.1158$\pm$0.0115 & G \\
2018Jan07 & 2018.0170  & Sa,Sb &    87.54  $\pm$   2.97  &     8.54  $\pm$   1.94  &  Kcon & 0.1178$\pm$0.0081 & G \\
          &            &       &                         &                         &  BrA  & 0.0597$\pm$0.0093 & G \\
2018Jan12 & 2018.0310  & Sa,Sb &    86.91  $\pm$   3.38  &     9.17  $\pm$   2.23  &  Kcon & 0.1073$\pm$0.0072 & G \\
          &            &       &                         &                         &   BrA & 0.0382$\pm$0.0050 & G \\
2018Jan18 & 2018.0471  & Sa,Sb &    88.50  $\pm$   5.29  &     8.72  $\pm$   3.42  &  Kcon & 0.1391$\pm$0.0093 & G \\
          &            &       &                         &                         &   BrA & 0.0548$\pm$0.0050 & G \\
2018Feb12 & 2018.1158  & Sa,Sb &    87.58  $\pm$   2.69  &     9.03  $\pm$   1.76  &  Kcon & 0.1714$\pm$0.0070 & G \\
          &            &       &                         &                         &   BrA & 0.0633$\pm$0.0065 & G \\
2018Feb13 & 2018.1186  & Sa,Sb &    86.21  $\pm$   3.40  &     9.34  $\pm$   2.26  &  Kcon & 0.1434$\pm$0.0069 & G \\
          &            &       &                         &                         &   BrA & 0.0588$\pm$0.0100 & G \\
2018Nov06 & 2018.8475  & Sa,Sb &    89.56  $\pm$  17.69  &    12.88  $\pm$  11.32  &  Kcon & 0.2848$\pm$0.2446 & G \\
          &            &       &                         &                         &   BrA & 0.0277$\pm$0.0120 & G \\
2018Nov12 & 2018.8637  & Sa,Sb &    72.23  $\pm$   4.41  &    16.55  $\pm$   3.50  &  Kcon & 0.2189$\pm$0.0266 & G \\
          &            &       &                         &                         &   BrA & 0.0404$\pm$0.0141 & G \\
          &            &       &                         &                         &  Hcon & 0.9231$\pm$0.5198 & G \\
2018Nov29 & 2018.9103  & Sa,Sb &    79.20  $\pm$   1.89  &    18.42  $\pm$   1.37  &  Kcon & 0.1543$\pm$0.0071 & G \\
          &            &       &                         &                         &   BrA & 0.0447$\pm$0.0094 & G \\
          &            &       &                         &                         &  Hcon & 0.6061$\pm$0.2147 & G \\
2018Dec02 & 2018.9186  & Sa,Sb &    80.03  $\pm$   1.89  &    18.34  $\pm$   1.35  &  Kcon & 0.1762$\pm$0.0097 & G \\
          &            &       &                         &                         &   BrA & 0.0514$\pm$0.0060 & G \\
          &            &       &                         &                         &  Hcon & 0.8400$\pm$0.2729 & G \\
2018Dec22 & 2018.9731  & Sa,Sb &    78.11  $\pm$   3.99  &    15.08  $\pm$   2.93  &  Kcon & 0.1531$\pm$0.0120 & G \\
          &            &       &                         &                         &   BrA & 0.0487$\pm$0.0080 & G \\
          &            &       &                         &                         &  Hcon & 0.4500$\pm$0.2636 & G \\
2019Jan02 & 2019.0032  & Sa,Sb &    74.52  $\pm$   2.97  &    17.16  $\pm$   2.28  &  Kcon & 0.1078$\pm$0.0116 & G \\
          &            &       &                         &                         &   BrA & 0.0596$\pm$0.0276 & G \\
          &            &       &                         &                         &  Hcon & 0.3115$\pm$0.0747 & G \\
2019Jan16 & 2019.0410  & Sa,Sb &    65.29  $\pm$  11.23  &    18.88  $\pm$   9.86  &  Kcon & 0.0900$\pm$0.0705 & G \\
          &            &       &                         &                         &   BrA & 0.0677$\pm$0.0125 & G \\
2019Jan20 & 2019.0520  & Sa,Sb &    78.34  $\pm$   1.76  &    20.04  $\pm$   1.29  & Kcont & 0.0438$\pm$0.0057 & K \\
          &            &       &                         &                         & Hcont & 0.0816$\pm$0.0693 & K \\
          &            &       &                         &                         &   PAH & 0.0210$\pm$0.0057 & K \\
\hline
2015Jan01 & 2015.0000  & N,Sa  &   689.91  $\pm$   0.72  &  191.524  $\pm$  0.061  & Kcont & 0.1353$\pm$0.0021 & K \\
          &            &       &                         &                         & Hcont & 0.0038$\pm$0.0006 & K \\
2015Apr05 & 2015.2573  & N,Sa  &   689.17  $\pm$   0.43  &  191.721  $\pm$  0.037  & Kcont & 0.1768$\pm$0.0011 & K \\
2016Oct20 & 2016.8021  & N,Sa  &   685.34  $\pm$   0.53  &  193.179  $\pm$  0.048  & Kcont & 0.2631$\pm$0.0035 & K \\
          &            &       &                         &                         & Hcont & 0.0159$\pm$0.0006 & K \\
2017Oct05 & 2017.7605  & N,Sa  &   688.19  $\pm$   0.98  &  194.151  $\pm$  0.096  &  Kcon & 0.1226$\pm$0.0026 & G \\
          &            &       &                         &                         &   BrA & 1.1090$\pm$0.0186 & G \\
2017Oct19 & 2017.7989  & N,Sa  &   688.79  $\pm$   1.16  &  194.201  $\pm$  0.108  &  Kcon & 0.2315$\pm$0.0030 & G \\
          &            &       &                         &                         &   BrA & 1.1723$\pm$0.0083 & G \\
2017Nov06 & 2017.8479  & N,Sa  &   688.00  $\pm$   0.86  &  194.221  $\pm$  0.087  &  Kcon & 0.1765$\pm$0.0023 & G \\
          &            &       &                         &                         &   BrA & 0.9774$\pm$0.0103 & G \\
2017Dec09 & 2017.9382  & N,Sa  &   687.81  $\pm$   0.81  &  194.325  $\pm$  0.084  &  Kcon & 0.1099$\pm$0.0020 & G \\
          &            &       &                         &                         &   BrA & 0.8732$\pm$0.0080 & G \\
2017Dec24 & 2017.9792  & N,Sa  &   687.74  $\pm$   0.80  &  194.348  $\pm$  0.083  &  Kcon & 0.1027$\pm$0.0016 & G \\
          &            &       &                         &                         &   BrA & 0.8629$\pm$0.0053 & G \\
2017Dec26 & 2017.9848  & N,Sa  &   687.31  $\pm$   1.25  &  194.376  $\pm$  0.115  &  Kcon & 0.0852$\pm$0.0022 & G \\
          &            &       &                         &                         &   BrA & 0.7487$\pm$0.0073 & G \\
2017Dec31 & 2017.9983  & N,Sa  &   687.39  $\pm$   1.18  &  194.374  $\pm$  0.110  &  Kcon & 0.0950$\pm$0.0015 & G \\
2018Jan05 & 2018.0120  & N,Sa  &   687.40  $\pm$   0.97  &  194.375  $\pm$  0.095  &  Kcon & 0.1185$\pm$0.0020 & G \\
2018Jan07 & 2018.0170  & N,Sa  &   687.35  $\pm$   0.92  &  194.369  $\pm$  0.091  &  Kcon & 0.1145$\pm$0.0026 & G \\
          &            &       &                         &                         &  BrA  & 0.9198$\pm$0.0062 & G \\
2018Jan12 & 2018.0310  & N,Sa  &   687.55  $\pm$   0.82  &  194.410  $\pm$  0.085  &  Kcon & 0.1253$\pm$0.0020 & G \\
          &            &       &                         &                         &   BrA & 1.0181$\pm$0.0058 & G \\
2018Jan18 & 2018.0471  & N,Sa  &   687.63  $\pm$   0.92  &  194.391  $\pm$  0.091  &  Kcon & 0.0921$\pm$0.0030 & G \\
          &            &       &                         &                         &   BrA & 0.8120$\pm$0.0080 & G \\
2018Feb12 & 2018.1158  & N,Sa  &   687.40  $\pm$   0.84  &  194.483  $\pm$  0.086  &  Kcon & 0.0689$\pm$0.0014 & G \\
          &            &       &                         &                         &   BrA & 0.6342$\pm$0.0058 & G \\
2018Feb13 & 2018.1186  & N,Sa  &   687.44  $\pm$   0.95  &  194.471  $\pm$  0.093  &  Kcon & 0.0828$\pm$0.0014 & G \\
          &            &       &                         &                         &   BrA & 0.7490$\pm$0.0114 & G \\
2018Nov06 & 2018.8475  & N,Sa  &   691.88  $\pm$  12.82  &  195.159  $\pm$  1.063  &  Kcon & 0.0489$\pm$0.0062 & G \\
          &            &       &                         &                         &   BrA & 0.7954$\pm$0.0087 & G \\
2018Nov12 & 2018.8637  & N,Sa  &   689.04  $\pm$   1.40  &  195.171  $\pm$  0.127  &  Kcon & 0.0504$\pm$0.0012 & G \\
          &            &       &                         &                         &   BrA & 0.8019$\pm$0.0105 & G \\
          &            &       &                         &                         &  Hcon & 0.0026$\pm$0.0007 & G \\
2018Nov29 & 2018.9103  & N,Sa  &   685.37  $\pm$   0.88  &  194.866  $\pm$  0.089  &  Kcon & 0.0663$\pm$0.0011 & G \\
          &            &       &                         &                         &   BrA & 0.8334$\pm$0.0050 & G \\
          &            &       &                         &                         &  Hcon & 0.0033$\pm$0.0005 & G \\
2018Dec02 & 2018.9186  & N,Sa  &   684.98  $\pm$   0.99  &  194.871  $\pm$  0.097  &  Kcon & 0.0593$\pm$0.0011 & G \\
          &            &       &                         &                         &   BrA & 0.8275$\pm$0.0055 & G \\
          &            &       &                         &                         &  Hcon & 0.0025$\pm$0.0005 & G \\
2018Dec22 & 2018.9731  & N,Sa  &   685.82  $\pm$   1.14  &  195.335  $\pm$  0.107  &  Kcon & 0.0764$\pm$0.0012 & G \\
          &            &       &                         &                         &   BrA & 0.9663$\pm$0.0047 & G \\
          &            &       &                         &                         &  Hcon & 0.0040$\pm$0.0008 & G \\
2019Jan02 & 2019.0032  & N,Sa  &   685.70  $\pm$   0.92  &  195.351  $\pm$  0.091  &  Kcon & 0.0985$\pm$0.0014 & G \\
          &            &       &                         &                         &   BrA & 1.0380$\pm$0.0155 & G \\
          &            &       &                         &                         &  Hcon & 0.0061$\pm$0.0004 & G \\
2019Jan16 & 2019.0410  & N,Sa  &   685.79  $\pm$   1.81  &  195.364  $\pm$  0.159  &  Kcon & 0.0852$\pm$0.0045 & G \\
          &            &       &                         &                         &   BrA & 0.8511$\pm$0.0114 & G \\
2019Jan20 & 2019.0520  & N,Sa  &   678.29  $\pm$   0.44  &  195.305  $\pm$  0.042  & Kcont & 0.1294$\pm$0.0014 & K \\
          &            &       &                         &                         & Hcont & 0.0049$\pm$0.0003 & K \\
          &            &       &                         &                         &   PAH & 0.4007$\pm$0.0065 & K \\
\hline
2015Jan01 & 2015.0000  & N,Sb  &   592.72  $\pm$   0.78  &  196.205  $\pm$  0.076  & Kcont & 0.0472$\pm$0.0009 & K \\
          &            &       &                         &                         & Hcont & 0.0109$\pm$0.0009 & K \\
2015Apr05 & 2015.2573  & N,Sb  &   592.76  $\pm$   0.61  &  196.105  $\pm$  0.060  & Kcont & 0.0639$\pm$0.0008 & K \\
2016Oct20 & 2016.8021  & N,Sb  &   592.53  $\pm$   2.31  &  195.650  $\pm$  0.224  & Kcont & 0.0202$\pm$0.0018 & K \\
          &            &       &                         &                         & Hcont & 0.0025$\pm$0.0006 & K \\
2017Oct05 & 2017.7605  & N,Sb  &   597.14  $\pm$   6.44  &  195.257  $\pm$  0.619  &  Kcon & 0.0190$\pm$0.0013 & G \\
          &            &       &                         &                         &   BrA & 0.0912$\pm$0.0204 & G \\
2017Oct19 & 2017.7989  & N,Sb  &   598.21  $\pm$   3.47  &  195.415  $\pm$  0.336  &  Kcon & 0.0239$\pm$0.0020 & G \\
          &            &       &                         &                         &   BrA & 0.0617$\pm$0.0092 & G \\
2017Nov06 & 2017.8479  & N,Sb  &   596.86  $\pm$   2.69  &  195.345  $\pm$  0.263  &  Kcon & 0.0205$\pm$0.0012 & G \\
          &            &       &                         &                         &   BrA & 0.0566$\pm$0.0088 & G \\
2017Dec09 & 2017.9382  & N,Sb  &   598.30  $\pm$   2.80  &  195.351  $\pm$  0.273  &  Kcon & 0.0126$\pm$0.0009 & G \\
          &            &       &                         &                         &   BrA & 0.0429$\pm$0.0070 & G \\
2017Dec24 & 2017.9792  & N,Sb  &   598.83  $\pm$   4.24  &  195.184  $\pm$  0.409  &  Kcon & 0.0136$\pm$0.0009 & G \\
          &            &       &                         &                         &   BrA & 0.0473$\pm$0.0070 & G \\
2017Dec26 & 2017.9848  & N,Sb  &   598.43  $\pm$   3.46  &  195.196  $\pm$  0.335  &  Kcon & 0.0131$\pm$0.0008 & G \\
          &            &       &                         &                         &   BrA & 0.0451$\pm$0.0057 & G \\
2017Dec31 & 2017.9983  & N,Sb  &   600.55  $\pm$   3.31  &  195.200  $\pm$  0.320  &  Kcon & 0.0120$\pm$0.0008 & G \\
2018Jan05 & 2018.0120  & N,Sb  &   600.21  $\pm$   3.23  &  195.250  $\pm$  0.312  &  Kcon & 0.0137$\pm$0.0012 & G \\
2018Jan07 & 2018.0170  & N,Sb  &   600.32  $\pm$   2.96  &  195.218  $\pm$  0.286  &  Kcon & 0.0135$\pm$0.0008 & G \\
          &            &       &                         &                         &  BrA  & 0.0549$\pm$0.0084 & G \\
2018Jan12 & 2018.0310  & N,Sb  &   601.06  $\pm$   3.36  &  195.166  $\pm$  0.324  &  Kcon & 0.0134$\pm$0.0008 & G \\
          &            &       &                         &                         &   BrA & 0.0389$\pm$0.0050 & G \\
2018Jan18 & 2018.0471  & N,Sb  &   599.63  $\pm$   5.05  &  195.226  $\pm$  0.485  &  Kcon & 0.0128$\pm$0.0006 & G \\
          &            &       &                         &                         &   BrA & 0.0445$\pm$0.0038 & G \\
2018Feb12 & 2018.1158  & N,Sb  &   600.27  $\pm$   2.83  &  195.278  $\pm$  0.275  &  Kcon & 0.0118$\pm$0.0004 & G \\
          &            &       &                         &                         &   BrA & 0.0401$\pm$0.0038 & G \\
2018Feb13 & 2018.1186  & N,Sb  &   601.62  $\pm$   3.46  &  195.206  $\pm$  0.333  &  Kcon & 0.0119$\pm$0.0005 & G \\
          &            &       &                         &                         &   BrA & 0.0440$\pm$0.0068 & G \\
2018Nov06 & 2018.8475  & N,Sb  &   602.40  $\pm$  23.61  &  195.497  $\pm$  2.246  &  Kcon & 0.0126$\pm$0.0066 & G \\
          &            &       &                         &                         &   BrA & 0.0220$\pm$0.0095 & G \\
2018Nov12 & 2018.8637  & N,Sb  &   616.83  $\pm$   4.84  &  195.010  $\pm$  0.452  &  Kcon & 0.0110$\pm$0.0011 & G \\
          &            &       &                         &                         &   BrA & 0.0323$\pm$0.0110 & G \\
          &            &       &                         &                         &  Hcon & 0.0024$\pm$0.0006 & G \\
2018Nov29 & 2018.9103  & N,Sb  &   606.35  $\pm$   2.17  &  194.402  $\pm$  0.211  &  Kcon & 0.0102$\pm$0.0004 & G \\
          &            &       &                         &                         &   BrA & 0.0372$\pm$0.0077 & G \\
          &            &       &                         &                         &  Hcon & 0.0020$\pm$0.0006 & G \\
2018Dec02 & 2018.9186  & N,Sb  &   605.12  $\pm$   2.38  &  194.413  $\pm$  0.231  &  Kcon & 0.0104$\pm$0.0004 & G \\
          &            &       &                         &                         &   BrA & 0.0425$\pm$0.0048 & G \\
          &            &       &                         &                         &  Hcon & 0.0021$\pm$0.0005 & G \\
2018Dec22 & 2018.9731  & N,Sb  &   607.71  $\pm$   3.84  &  195.368  $\pm$  0.365  &  Kcon & 0.0117$\pm$0.0009 & G \\
          &            &       &                         &                         &   BrA & 0.0470$\pm$0.0077 & G \\
          &            &       &                         &                         &  Hcon & 0.0018$\pm$0.0006 & G \\
2019Jan02 & 2019.0032  & N,Sb  &   611.22  $\pm$   2.98  &  195.131  $\pm$  0.284  &  Kcon & 0.0106$\pm$0.0011 & G \\
          &            &       &                         &                         &   BrA & 0.0615$\pm$0.0271 & G \\
          &            &       &                         &                         &  Hcon & 0.0019$\pm$0.0005 & G \\
2019Jan16 & 2019.0410  & N,Sb  &   620.63  $\pm$  12.43  &  194.994  $\pm$  1.149  &  Kcon & 0.0074$\pm$0.0045 & G \\
          &            &       &                         &                         &   BrA & 0.0575$\pm$0.0101 & G \\
2019Jan20 & 2019.0520  & N,Sb  &   600.26  $\pm$   1.75  &  194.688  $\pm$  0.168  & Kcont & 0.0057$\pm$0.0007 & K \\
          &            &       &                         &                         & Hcont & 0.0004$\pm$0.0003 & K \\
          &            &       &                         &                         &   PAH & 0.0084$\pm$0.0022 & K \\
\enddata 
\tablenotetext{a}{The last column identifies the telescope used: G = Gemini North, K = Keck II.} 
\end{deluxetable} 


\begin{deluxetable}{lcccc}    
\tablewidth{0pt}
\tablecaption{Absolute photometry based on Gemini observations of T Tau. \label{tab.phot}} 
\tablehead{
\colhead{JY} & \colhead{Filter} & \colhead{N} & \colhead{Sa} & \colhead{Sb}}
\startdata 
2017.8479 & $K$ & 5.52 $\pm$ 0.04 &  7.40 $\pm$ 0.08 &  9.74 $\pm$ 0.33 \\
2017.9792 & $K$ & 5.51 $\pm$ 0.03 &  7.98 $\pm$ 0.09 & 10.17 $\pm$ 0.39 \\
2017.9848 & $K$ & 5.46 $\pm$ 0.06 &  8.13 $\pm$ 0.16 & 10.16 $\pm$ 0.37 \\
2018.0120 & $K$ & 5.60 $\pm$ 0.03 &  7.91 $\pm$ 0.10 & 10.26 $\pm$ 0.50 \\
2018.0170 & $K$ & 5.59 $\pm$ 0.03 &  7.94 $\pm$ 0.13 & 10.26 $\pm$ 0.34 \\
2018.0310 & $K$ & 5.50 $\pm$ 0.03 &  7.75 $\pm$ 0.10 & 10.17 $\pm$ 0.34 \\
2018.0471 & $K$ & 5.46 $\pm$ 0.04 &  8.05 $\pm$ 0.19 & 10.19 $\pm$ 0.26 \\
2018.1158 & $K$ & 5.45 $\pm$ 0.06 &  8.36 $\pm$ 0.13 & 10.27 $\pm$ 0.20 \\
2018.1186 & $K$ & 5.50 $\pm$ 0.03 &  8.20 $\pm$ 0.10 & 10.31 $\pm$ 0.24 \\
2018.8475 & $K$ & 5.50 $\pm$ 0.04 &  8.78 $\pm$ 0.72 & 10.25 $\pm$ 3.00 \\
2018.8637 & $K$ & 5.53 $\pm$ 0.05 &  8.77 $\pm$ 0.15 & 10.43 $\pm$ 0.58 \\
2018.9103 & $K$ & 5.60 $\pm$ 0.07 &  8.54 $\pm$ 0.12 & 10.57 $\pm$ 0.23 \\
2018.9186 & $K$ & 5.63 $\pm$ 0.03 &  8.69 $\pm$ 0.11 & 10.58 $\pm$ 0.24 \\
2018.9731 & $K$ & 5.65 $\pm$ 0.03 &  8.44 $\pm$ 0.09 & 10.48 $\pm$ 0.42 \\
2019.0032 & $K$ & 5.64 $\pm$ 0.07 &  8.16 $\pm$ 0.10 & 10.57 $\pm$ 0.59 \\
2019.0410 & $K$ & 5.57 $\pm$ 0.03 &  8.24 $\pm$ 0.30 & 10.90 $\pm$ 3.50 \\
2017.8479 & $L$ & 5.79 $\pm$ 0.33 &  5.81 $\pm$ 0.33 &  8.90 $\pm$ 0.94 \\
2017.9792 & $L$ & 5.17 $\pm$ 0.60 &  5.33 $\pm$ 0.60 &  8.48 $\pm$ 1.04 \\
2017.9848 & $L$ & 5.50 $\pm$ 0.32 &  5.81 $\pm$ 0.32 &  8.86 $\pm$ 0.80 \\
2018.0170 & $L$ & 5.58 $\pm$ 0.66 &  5.67 $\pm$ 0.66 &  8.73 $\pm$ 1.10 \\
2018.0310 & $L$ & 5.02 $\pm$ 0.64 &  5.00 $\pm$ 0.64 &  8.54 $\pm$ 0.98 \\
2018.0471 & $L$ & 5.52 $\pm$ 0.46 &  5.75 $\pm$ 0.46 &  8.90 $\pm$ 0.67 \\
2018.1158 & $L$ & 5.36 $\pm$ 0.53 &  5.86 $\pm$ 0.53 &  8.85 $\pm$ 0.76 \\
2018.1186 & $L$ & 5.23 $\pm$ 0.45 &  5.54 $\pm$ 0.46 &  8.62 $\pm$ 1.00 \\
2018.8637 & $L$ & 5.53 $\pm$ 1.83 &  5.77 $\pm$ 1.83 &  9.26 $\pm$ 2.68 \\
2018.9103 & $L$ & 5.03 $\pm$ 0.87 &  5.23 $\pm$ 0.87 &  8.60 $\pm$ 1.47 \\
2018.9186 & $L$ & 5.49 $\pm$ 0.58 &  5.70 $\pm$ 0.58 &  8.92 $\pm$ 0.87 \\
2018.9731 & $L$ & 5.10 $\pm$ 0.23 &  5.14 $\pm$ 0.24 &  8.42 $\pm$ 0.96 \\
2019.0032 & $L$ & 5.34 $\pm$ 1.17 &  5.30 $\pm$ 1.17 &  8.36 $\pm$ 2.78 \\
2019.0410 & $L$ & 5.26 $\pm$ 0.65 &  5.43 $\pm$ 0.65 &  8.36 $\pm$ 1.19 \\
2018.8637 & $H$ & 6.37 $\pm$ 0.03 & 12.83 $\pm$ 1.59 & 12.92 $\pm$ 1.37 \\
2018.9103 & $H$ & 6.36 $\pm$ 0.04 & 12.56 $\pm$ 0.82 & 13.11 $\pm$ 1.64 \\
2018.9186 & $H$ & 6.40 $\pm$ 0.06 & 12.91 $\pm$ 1.13 & 13.09 $\pm$ 1.34 \\
2018.9731 & $H$ & 6.44 $\pm$ 0.04 & 12.44 $\pm$ 1.12 & 13.30 $\pm$ 1.95 \\
2019.0032 & $H$ & 6.41 $\pm$ 0.04 & 11.94 $\pm$ 0.34 & 13.21 $\pm$ 1.51 \\
\enddata 
\end{deluxetable} 

\clearpage

\begin{deluxetable}{lc}    
\tablewidth{0pt}
\tablecaption{Orbital parameters of T Tau Sa,Sb \label{tab.orb}} 
\tablehead{
\colhead{Parameter}      & \colhead{Value}}
\startdata 
$P$ (yr)              &   27.18   $\pm$ 0.72      \\
$T$ (JY)             &   1996.10 $\pm$ 0.38      \\
$e$                  &   0.551   $\pm$ 0.032     \\
$a$ (mas)            &   85.12   $\pm$ 0.62      \\
$i$ ($^\circ$)        &  21.1    $\pm$ 2.1         \\
$\Omega$ ($^\circ$)   &  94.4    $\pm$ 16.9       \\
$\omega$ ($^\circ$)   &  45.8    $\pm$ 16.9        \\
$M_{\rm Sb}/M_{\rm Sa}$ &  0.210   $\pm$ 0.028  \\
\enddata
\end{deluxetable} 

\begin{deluxetable}{lcc}    
\tablewidth{0pt}
\tablecaption{Dynamical Masses of T Tau Sa and Sb \label{tab.mass}} 
\tablehead{
\colhead{Parameter}      & \colhead{VLBA Parallax}  & \colhead{Gaia Parallax}}
\startdata          
Adopted $d$ (pc)    & 148.7 $\pm$ 1.0    & 143.74 $\pm$ 1.22     \\  
Reference           & \citet{galli18}    & \citet{bailerjones18} \\
$M_{\rm Sa+Sb} (M_\odot)$ & 2.744 $\pm$ 0.166 & 2.479 $\pm$ 0.155 \\
$M_{\rm Sa} (M_\odot)$ & 2.268 $\pm$  0.147 & 2.049 $\pm$ 0.137     \\  
$M_{\rm Sb} (M_\odot)$ & 0.476 $\pm$ 0.060  & 0.430 $\pm$ 0.055     \\  
\enddata
\end{deluxetable} 

\begin{deluxetable}{lcc}    
\tablewidth{0pt}
\tablecaption{Range of Orbital Parameters for T Tau N,S \label{tab.wide}} 
\tablehead{
\colhead{Parameter}      & \colhead{Best Fit}  & \colhead{Range} }
\startdata 
$P$ (yr)              &   4602.6  &  481 $-$ 4997 \\
$T$ (JY)             &   1951.3  & 1697 $-$ 2344 \\
$e$                  &   0.754   & 0.00 $-$ 0.79 \\
$a$ (mas)            &   3255.1  & 733 $-$ 3426  \\
$i$ ($^\circ$)        &    54.2   & 29 $-$   60  \\
$\Omega$ ($^\circ$)   &    148.2  & 70 $-$  164  \\
$\omega$ ($^\circ$)   &    10.4   &  0 $-$  360   \\
\enddata
\tablecomments{We applied a constraint of 4.51 $\pm$ 0.59 $M_\odot$ on the total mass of N+Sa+Sb and an upper limit on the period of 5000 yr.}
\end{deluxetable}


\end{document}